\documentclass[aps,11pt,article]{revtex4}
\usepackage{times}
\usepackage{amssymb}
\usepackage{amsmath}
\usepackage{graphicx}
\usepackage{epsfig}
\usepackage{dcolumn}
\usepackage{color}
\usepackage{bm}
\begin{document}
\title{Quantum theory of nonlinear electromagnetic response}

\author{Anwei Zhang}
\email{zhanganwei@shnu.edu.cn}
\affiliation{Department of Physics, Mathematics $\&$ Science College, Shanghai Normal University, No. 100 Guilin Road, Shanghai, 200234 China}
\author{C. M. Wang}
\affiliation{Department of Physics, Mathematics $\&$ Science College, Shanghai Normal University, No. 100 Guilin Road, Shanghai, 200234 China}

\begin{abstract}
In recent years, the investigation  of nonlinear electromagnetic responses has received significant attention  due to  its potential for elucidating the quantum properties of matter. Although  remarkable  progress has been achieved in developing quantum theories of nonlinear responses to electric field, a comprehensive quantum theory framework that systematically addresses nonlinear responses to both electric and magnetic fields has yet to be thoroughly discussed.
Here, we present a systematic quantum theory of nonlinear electromagnetic response using the Matsubara Green's function approach, which explicitly incorporates the wave vector dependence of external electromagnetic fields. 
We reveal the general properties of transport coefficients. We apply our theory to second-order responses, deriving  the nonlinear Hall effects and magneto-nonlinear Hall effects in both time-reversal symmetric and time-reversal breaking systems. These effects stem from diverse  quantum geometric quantities. Additionally, we analyze  the contributions arising  from the Zeeman interaction.
Our work presents  a unified quantum theory of nonlinear electromagnetic response, 
paving the way for further exploration of novel phenomena in this field.
\end{abstract}

\maketitle

\section{Introduction}
The electromagnetic response describes how a condensed-matter system reacts to an external electromagnetic field, typically manifested as an electric current. It is a crucial tool for investigating condensed-matter physics and can reveal many key properties of materials. In particular, the nonlinear electromagnetic response provides a unique perspective for understanding the dynamic mechanisms~\cite{sipe2000second,cook2017design,parker2019diagrammatic,ahn2020low,du2021quantum,kaplan2023general}, structural symmetry~\cite{gao2014field,sodemann2015quantum,ma2019observation,wang2024orbital,zhang2023symmetry}, and quantum geometry ~\cite{morimoto2016topological,liu2021intrinsic,bhalla2022resonant,gao2023quantum,zhai2023large,zheng2024interlayer} of materials. Over the past few years, nonlinear electromagnetic response has rapidly developed in the interdisciplinary field of quantum optics and condensed-matter physics, demonstrating significant research potential and application value~\cite{du2021nonlinear,ma2021topology,10.1093/nsr/nwae334}. For example, it provides a theoretical foundation for designing efficient optoelectronic and quantum devices.

The theoretical study of the nonlinear electromagnetic response relies on both semi-classical and quantum approaches. Semi-classical methods, such as the semi-classical equations of motion for electron wave packet ~\cite{gao2014field} and the Boltzmann equation ~\cite{sodemann2015quantum}, are well-suited for describing macroscopic electron transport behavior under weak disorder and low-energy excitation. However, these methods have clear limitations  when dealing with strong disorder and nonlinear effects ~\cite{du2019disorder}. In contrast, quantum methods, such as 
the Floquet formalism~\cite{morimoto2016semiclassical}, density matrix method~\cite{gao2020second}, and Feynman diagrammatic technique ~\cite{parker2019diagrammatic,du2021nonlinear}, can fully capture the quantum behavior of electrons, thereby offering more accurate descriptions, especially in the presence of strong disorder and nonlinear effects.
Despite these advances, the current quantum theories of nonlinear responses are primarily confined to the field of electric field driving, and the quantum
theory of nonlinear response involving magnetic field is still incomplete. Most studies of nonlinear responses involving magnetic fields are  based on the semi-classical equations of motion for Bloch electrons~\cite{gao2014field,morimoto2016semiclassical}, which are insufficient for an in-depth exploration of material properties. Given this gap, there is an urgent need to construct a systematic quantum theory of the nonlinear response to the input electromagnetic field.

In this paper, we develop a comprehensive quantum theoretical framework for nonlinear electromagnetic responses by employing the Matsubara Green’s function method, which incorporates the wave vector of the external electromagnetic field. This approach offers several key advantages when addressing quantum many-body systems at finite temperatures. Specifically, it naturally accommodates finite-temperature effects, seamlessly integrates with Feynman diagram techniques, and effectively handles complex interactions.
Within this framework, we 
reveal universal constraints governing the transport coefficients for various  types of electromagnetic responses.
Furthermore, we derive the nonlinear Hall effects in both time-reversal symmetric systems with Berry curvature dipole structures and time-reversal breaking systems with normalized quantum metric dipole structures. Our analysis covers the general case as well as the specific scenario of second-harmonic generation (SHG). This work elucidates the diverse mechanisms underlying these phenomena in systems with distinct symmetry properties and reveals how they manifest under different DC limits.  We also introduce two distinct types of magneto-nonlinear Hall effects that emerge in time-reversal symmetric and time-reversal breaking systems, respectively. These findings  enrich the theoretical understanding of nonlinear electromagnetic phenomena. Additionally, we  examine the contributions from the Zeeman interaction.
Overall, this work establishes a robust theoretical foundation for investigating nonlinear electromagnetic responses. It is expected to stimulate further research in this field and contribute to the development of advanced quantum devices with precisely engineered electromagnetic properties.



\section{Setup}
We consider a periodic crystalline system which is described by the Hamiltonian $H_0$. According the Bloch’s theorem, the Hamiltonian $H_0$ can be diagonalized as $H_0|\psi_a(\mathbf{k})\rangle=\varepsilon_a(\mathbf{k})|\psi_a(\mathbf{k})\rangle$, where $|\psi_a(\mathbf{k})\rangle=e^{i\mathbf{k}\cdot \mathbf{r}}|u_a(\mathbf{k})\rangle$ is the Bloch wave function labeled by band index $a$ and Bloch wave vector $\mathbf{k}$. Here, $|u_a(\mathbf{k})\rangle$ represents the periodic part of the Bloch wave  and $\varepsilon_a(\mathbf{k})$ denotes the energy dispersion of band $a$.

In the presence of an external electromagnetic field,  the minimal substitution scheme $\mathbf{p} \rightarrow \mathbf{p}+e \mathbf{A}$ gives rise to the total Hamiltonian $H$, where $\mathbf{p}$ is the canonical momentum of the system, $-e$ is the charge of the electron, and $\mathbf{A}$ is the vector potential. To study the linear and nonlinear properties of optical response, we expand the total Hamiltonian  in terms of the vector potential as Taylor series
\begin{equation}\label{e1}
 H=H_0+H^{'}=H_0+e \mathbf{A} \cdot \partial_{\mathbf{p}}H_0+\frac{1}{2}e^2 \mathbf{A} \cdot \partial_{\mathbf{p}}(\mathbf{A} \cdot \partial_{\mathbf{p}}H_0)+\cdots,
\end{equation}
where $H^{'}$ is the perturbed Hamiltonian due to the external field.
Since $\partial_{\mathbf{p}}H_0$ is just the velocity operator $\mathbf{\hat{v}}$ in the absence of the  field, the interaction Hamiltonian can be rewritten as
\begin{equation}\label{e2}
 H^{'}=\frac{e}{2} (\mathbf{A} \cdot \mathbf{\hat{v}}+\mathbf{\hat{v}} \cdot \mathbf{A})+\frac{e^2}{8} \big[\mathbf{A} \cdot \partial_{\mathbf{p}}(\mathbf{A} \cdot \mathbf{\hat{v}})+\mathbf{A} \cdot \partial_{\mathbf{p}}(\mathbf{\hat{v}} \cdot \mathbf{A})+ \partial_{\mathbf{p}}(\mathbf{A} \cdot \mathbf{\hat{v}})\cdot \mathbf{A}+\partial_{\mathbf{p}}(\mathbf{\hat{v}}\cdot\mathbf{A})\cdot \mathbf{A}\big]+\cdots.
\end{equation}
Here, the interaction Hamiltonian has been symmetrized and we have set $\hbar=1$.
The total velocity operator in the presence of the electromagnetic field is given by
\begin{equation}\label{e3}
\mathbf{\hat{v}}^{tot}=\partial_{\mathbf{p}}H=\mathbf{\hat{v}}+\partial_{\mathbf{p}}H^{'}.
\end{equation}
The current density operator in momentum space  is expressed as~\cite{dressel2002electrodynamics,zhong2016gyrotropic}
\begin{equation}\label{e4}
\mathbf{\hat{j}}(\mathbf{q})=-\frac{e}{2V}(\mathbf{\hat{v}}^{tot}e^{-i\mathbf{q}\cdot \mathbf{r}}+e^{-i\mathbf{q}\cdot \mathbf{r}}\mathbf{\hat{v}}^{tot}),
\end{equation}
where $\mathbf{q}$ is the wave vector
and $V$ is the volume of the system.

\section{average value of current operator}

Next, we explore the average value of the current density operator, which can be  expressed through  Green's function. 
 Using  the second quantization method, the average value of the current density operator (a one-body operator) can be written as~\cite{jishi2013feynman}
\begin{equation}\label{01}
\langle \mathbf{\hat{j}}(\mathbf{q},\tau) \rangle=\sum_{n,m,\mathbf{k'},\mathbf{k}}\langle \psi_n(\mathbf{k'})|\mathbf{\hat{j}}(\mathbf{q})|\psi_m(\mathbf{k})\rangle\langle \hat{a}^\dag_{n}(\mathbf{k'},\tau)  \hat{a}_{m}(\mathbf{k},\tau)\rangle.
\end{equation}
Here, $\tau$ is the imaginary-time,  and $\hat{a}^\dag_{m}(\mathbf{k'},\tau)$ and $\hat{a}_{n}(\mathbf{k},\tau
)$ are fermion creation and annihilation operators, respectively. At finite temperature, the average  refers to the grand canonical ensemble average, i.e., $\langle \dots \rangle=Z^{-1}\mathrm{Tr}[e^{-\beta \widetilde{H}}\dots]$, where  $Z=\mathrm{Tr}[e^{-\beta \widetilde{H}}]$ is the grand canonical partition function, $\beta=1/k_B T$ is the inverse temperature, $\widetilde{H}=H-\mu \hat{N}$, $\mu$  is the chemical potential,  $\hat{N}$ is the number of particles operator, and $\mathrm{Tr}$ denotes the trace.
Note that $\langle \psi_n(\mathbf{k'})|$, $\mathbf{\hat{j}}(\mathbf{q})$, and 
$|\psi_m(\mathbf{k})\rangle$ 
contain  exponential  factors $e^{-i\mathbf{k'}\cdot \mathbf{r}}$,  $e^{-i\mathbf{q}\cdot \mathbf{r}}$, and
 $e^{i\mathbf{k}\cdot \mathbf{r}}$, respectively.
By replacing the variable $\mathbf{r}$ with $\mathbf{r}+\mathbf{R}$, where $\mathbf{R}$ is any lattice vector, we obtain
\begin{eqnarray}\label{ee}
 &&\sum_{n,m,\mathbf{k'},\mathbf{k}}\langle \psi_n(\mathbf{k'})|\mathbf{\hat{j}}(\mathbf{q})|\psi_m(\mathbf{k})\rangle\langle \hat{a}^{\dag}_{n}(\mathbf{k'},\tau)  \hat{a}_{m}(\mathbf{k},\tau)\rangle \nonumber\\&=&
\sum_{n,m,\mathbf{k'},\mathbf{k}}\langle \psi_n(\mathbf{k'})|\mathbf{\hat{j}}(\mathbf{q})|\psi_m(\mathbf{k})\rangle\langle \hat{a}^\dag_{n}(\mathbf{k'},\tau)  \hat{a}_{m}(\mathbf{k},\tau)\rangle e^{-i(\mathbf{k'}+\mathbf{q}-\mathbf{k})\cdot \mathbf{R}}.
\end{eqnarray}
Since Eq.~(\ref{ee}) holds true for every lattice vector $\mathbf{R}$, it means that $\mathbf{k'}=\mathbf{k}-\mathbf{q}+\mathbf{G}$~\cite{jishi2013feynman}, with $\mathbf{G}$ being a reciprocal lattice vector. The wave vectors $\mathbf{k}$ and $\mathbf{k'}$ should lie within the first Brillouin zone (FBZ). If $\mathbf{k}-\mathbf{q}$ does not belong to FBZ, $\mathbf{G}$  brings it back to the FBZ. If $\mathbf{k}-\mathbf{q}$ is already in the FBZ, $\mathbf{G}$ is zero.  For visible and infrared light,  the quantity $\mathbf{q}$ is much small compared to the size of Brillouin zone. As a result, $\mathbf{k}-\mathbf{q}$ generally remains within  FBZ. This implies that $\mathbf{G}$ is  zero, and thus $\mathbf{k'}=\mathbf{k}-\mathbf{q}$. Consequently, 
 Eq.~(\ref{01}) becomes
\begin{eqnarray}\label{001}
\langle \mathbf{\hat{j}}(\mathbf{q},\tau) \rangle&=&\sum_{n,m,\mathbf{k}}\langle \psi_n(\mathbf{k}-\mathbf{q})|\mathbf{\hat{j}}(\mathbf{q})|\psi_m(\mathbf{k})\rangle\langle \hat{a}^\dag_{n}(\mathbf{k}-\mathbf{q},\tau)  \hat{a}_{m}(\mathbf{k},\tau)\rangle\nonumber\\
&=&-\sum_{n,m,\mathbf{k}}\langle \psi_n(\mathbf{k}-\mathbf{q})|\mathbf{\hat{j}}(\mathbf{q})|\psi_m(\mathbf{k})\rangle\langle T\hat{a}_{m}(\mathbf{k},\tau)\hat{a}^\dag_{n}(\mathbf{k}-\mathbf{q},\tau^{+})\rangle\nonumber\\
&=&\sum_{n,m,\mathbf{k}}\langle \psi_n(\mathbf{k}-\mathbf{q})|\mathbf{\hat{j}}(\mathbf{q})|\psi_m(\mathbf{k})\rangle G_{mn}(\mathbf{k},\tau;\mathbf{k}-\mathbf{q},\tau^{+})\nonumber\\
&=&\sum_{\mathbf{k}}\mathrm{Tr}\big[\mathbf{\hat{j}}(\mathbf{q}) G(\mathbf{k},\tau;\mathbf{k}-\mathbf{q},\tau^{+})\big],
\end{eqnarray}
where $T$ is the time-ordering operator, $\tau^{+}=\tau+0^{+}$, and $G_{mn}(\mathbf{k},\tau;\mathbf{k}-\mathbf{q},\tau^{+})$
is the matrix element of Green's function.
The annihilation operator can be decomposed into a Fourier series as
\begin{equation}\label{02}
 \hat{a}_{m}(\mathbf{k},\tau)=\frac{1}{\beta}\sum_{i\omega_n}  e^{-i\omega_n\tau}\hat{a}_{m}(\mathbf{k},i\omega_n),
\end{equation}
where $\omega_n=(2n+1) \pi /\beta$ is the fermionic Matsubara frequency.
Thus,  Eq.~(\ref{001}) can be rewritten as
\begin{eqnarray}\label{002}
\langle \mathbf{\hat{j}}(\mathbf{q},\tau) \rangle
&=&\frac{1}{\beta^2}\sum_{\mathbf{k},i\omega_n,i\omega_{n'}}e^{-i\omega_n\tau+i\omega_{n'}\tau^{+}}\mathrm{Tr}\big[\mathbf{\hat{j}}(\mathbf{q}) G(\mathbf{k},i\omega_n;\mathbf{k}-\mathbf{q},i\omega_{n'})\big]\nonumber\\
&=&\frac{1}{\beta^2}\sum_{\mathbf{k},i\omega_n,i\omega}e^{-i\omega\tau}\mathrm{Tr}\big[\mathbf{\hat{j}}(\mathbf{q}) G(\mathbf{k},i\omega_n;\mathbf{k}-\mathbf{q},i\omega_n-i\omega)\big],
\end{eqnarray}
where  we let $\omega_n-\omega_{n'}=\omega$. The average value of the  current density operator can also be decomposed into series as
\begin{eqnarray}\label{003}
\langle \mathbf{\hat{j}}(\mathbf{q},\tau) \rangle
&=&\frac{1}{\beta}\sum_{i\omega}e^{-i\omega\tau}\langle \mathbf{\hat{j}}(\mathbf{q},i\omega) \rangle.
\end{eqnarray}
Combining Eq.~(\ref{002}) with Eq.~(\ref{003}), we obtain
\begin{eqnarray}\label{04}
\langle \mathbf{\hat{j}}(\mathbf{q},i\omega) \rangle=\frac{1}{\beta}\sum_{\mathbf{k},i\omega_n}\mathrm{Tr}\big[\mathbf{\hat{j}}(\mathbf{q}) G(\mathbf{k},i\omega_n;\mathbf{k}-\mathbf{q},i\omega_n-i\omega)\big].
\end{eqnarray}
Here,  $G(\mathbf{k},i\omega_n;\mathbf{k}-\mathbf{q},i\omega_n-i\omega)$ is the exact Green's function in momentum-frequency space.
For a normal crystalline system with free electron, the Hamiltonian $H_0$ and $H'$ are both  one-body operators. Therefore, 
 the exact Green's function can be expanded as a perturbation series ~\cite{fjaerestad2013introduction}
\begin{eqnarray}\label{05}
G(\mathbf{k},i\omega_n;\mathbf{k}-\mathbf{q},i\omega_n-i\omega)&=&G(\mathbf{k},i\omega_n)+G(\mathbf{k},i\omega_n)H^{'}G(\mathbf{k}-\mathbf{q},i\omega_n-i\omega)\nonumber\\&&+\big[G(\mathbf{k},i\omega_n)H^{'}G(\mathbf{k}-\mathbf{q}_1,i\omega_n-i\omega_1)H^{'}
 G(\mathbf{k}-\mathbf{q},i\omega_n-i\omega)\nonumber\\&&+(1\leftrightarrow 2)\big]+\dots,
\end{eqnarray}
where $G(\mathbf{k},i\omega_n)$ is the unperturbed Green's function and in clean systems it takes the form 
\begin{equation}\label{06}
G(\mathbf{k},i\omega_n)=\sum_a \frac{|\psi_a(\mathbf{k})\rangle\langle \psi_a(\mathbf{k})|}{i\omega_n+\mu-\varepsilon_a(\mathbf{k})},
\end{equation}
($\mathbf{q}_1$, $\mathbf{q}_2$, $\cdots$) and ($\omega_1$, $\omega_2$, $\cdots$) represent  the wave vectors and frequencies of the electromagnetic fields, respectively. The conservation of momentum and energy implies that  $\mathbf{q}_1+\mathbf{q}_2=\mathbf{q}$ and $\omega_1+\omega_2=\omega$ in the third term of  equation ~(\ref{05}).

The vector potential of the electromagnetic field can be expressed as
\begin{equation}\label{07}
 \mathbf{A}=\sum_{l} \mathbf{A}(\mathbf{q}_l,\omega_{l}) e^{i(\mathbf{q}_{l}\cdot \mathbf{r}-\omega_{l} t)}.
\end{equation}
For a monochromatic  field,  the vector potential simplifies to $\mathbf{A}=\mathbf{A} (\mathbf{q},\omega) e^{i(\mathbf{q}\cdot \mathbf{r}-\omega t)}$, where $\mathbf{q}$ and  $\omega$  are the  wave vector and frequency, respectively.
In the subsequent  analysis, the time-dependent exponential factors $e^{-i\omega_{l} t}$ in the vector potential will be omitted. This is because  the conductivity is defined as the coefficient of $e^{-i\omega t}$ that relates the induced current to the external fields \cite{marder2010condensed}.  Moreover, these factors merely reflect the conservation of energy, which is expressed as $\omega=\sum_{l}\omega_{l}$.

Next, we  express the matrix elements in  Eq.~(\ref{04})  in terms of the periodic part of the Bloch wave.
Using  Eq.~(\ref{e2}) and Eq.~(\ref{07}), we can obtain
the matrix elements of the interaction Hamiltonian 
\begin{eqnarray}\label{08}
\langle \psi_a(\mathbf{k})|H'|\psi_b(\mathbf{k}-\mathbf{q})\rangle&=&\langle u_a(\mathbf{k})|\frac{e}{2} \mathbf{A}(\mathbf{q},\omega) \cdot\big[ \mathbf{\hat{v}}(\mathbf{k}-\mathbf{q})+\mathbf{\hat{v}}(\mathbf{k})\big]+\bigg\{\frac{e^2}{8} \mathbf{A}(\mathbf{q}_1,\omega_1)\cdot\partial_{\mathbf{p}}\big\{\mathbf{A}(\mathbf{q}_2,\omega_2)\nonumber\\&& \cdot \big[\mathbf{\hat{v}}(\mathbf{k})+\mathbf{\hat{v}}(\mathbf{k}-\mathbf{q}_1)+\mathbf{\hat{v}}(\mathbf{k}-\mathbf{q}_2)+\mathbf{\hat{v}}(\mathbf{k}-\mathbf{q})\big]\big\}+(1\leftrightarrow 
2)\bigg\}|u_b(\mathbf{k}-\mathbf{q})\rangle\nonumber\\&&+\cdots\nonumber\\&\equiv&\langle u_a(\mathbf{k})|H^{'}(\mathbf{k},\mathbf{k}-\mathbf{q})|u_b(\mathbf{k}-\mathbf{q})\rangle,
\end{eqnarray}
\begin{eqnarray}\label{09}
\langle \psi_a(\mathbf{k})|H'|\psi_b(\mathbf{k}-\mathbf{q}_1)\rangle&=&\langle u_a(\mathbf{k})|\frac{e}{2} \mathbf{A}(\mathbf{q}_1,\omega_1) \cdot\big[ \mathbf{\hat{v}}(\mathbf{k}-\mathbf{q}_1)+\mathbf{\hat{v}}(\mathbf{k})\big]|u_b(\mathbf{k}-\mathbf{q}_1)\rangle\nonumber\\&\equiv&\langle u_a(\mathbf{k})|H^{'}(\mathbf{k},\mathbf{k}-\mathbf{q}_1)|u_b(\mathbf{k}-\mathbf{q}_1)\rangle,
\end{eqnarray}
and
\begin{eqnarray}\label{10}
\langle \psi_a(\mathbf{k}-\mathbf{q}_1))|H'|\psi_b(\mathbf{k}-\mathbf{q}_1-\mathbf{q}_2)\rangle&=&\langle u_a(\mathbf{k}-\mathbf{q}_1)|\frac{e}{2} \mathbf{A}(\mathbf{q}_2,\omega_2) \cdot\big[ \mathbf{\hat{v}}(\mathbf{k}-\mathbf{q})+\mathbf{\hat{v}}(\mathbf{k}-\mathbf{q}_1)\big]|u_b(\mathbf{k}-\mathbf{q})\rangle\nonumber\nonumber\\&\equiv&\langle u_a(\mathbf{k}-\mathbf{q}_1))|H^{'}(\mathbf{k}-\mathbf{q}_1,\mathbf{k}-\mathbf{q}_1-\mathbf{q}_2)|u_b(\mathbf{k}-\mathbf{q})\rangle,
\end{eqnarray}
where $\mathbf{\hat{v}}(\mathbf{k})=e^{-i\mathbf{k}\cdot \mathbf{r}}\mathbf{\hat{v}}e^{i\mathbf{k}\cdot \mathbf{r}}=\partial_{\mathbf{k}}H({\mathbf{k}})$ and
$H({\mathbf{k}})=e^{-i\mathbf{k}\cdot \mathbf{r}}H_0e^{i\mathbf{k}\cdot \mathbf{r}}$.
 It is worth noting that $H({\mathbf{k}})$  satisfies the eigenvalue equation $H({\mathbf{k}})|u_a(\mathbf{k})\rangle=\varepsilon_a({\mathbf{k}})|u_a(\mathbf{k})\rangle$,  $\partial_{\mathbf{p}}\mathbf{\hat{v}}(\mathbf{k})$ is actually $-i e^{-i\mathbf{k}\cdot \mathbf{r}}[\mathbf{r},\mathbf{\hat{v}}]e^{i\mathbf{k}\cdot \mathbf{r}}=\partial_{\mathbf{k}}\mathbf{\hat{v}}(\mathbf{k})$,  and the first term in Eq.~(\ref{08}) corresponds to the case of a monochromatic  field, while the second term refers to the case of a bichromatic field.
Similarly, by combining Eq.~(\ref{e4}) and Eq.~(\ref{07}), we can obtain
the matrix elements of the current operator
\begin{eqnarray}\label{110}
\langle \psi_a(\mathbf{k}-\mathbf{q}_1)|\mathbf{\hat{j}}(\mathbf{q})|\psi_b(\mathbf{k})\rangle&=&\langle u_a(\mathbf{k}-\mathbf{q}_1)|-\frac{e^2}{4V} \partial_{\mathbf{k}}\big\{\mathbf{A}(\mathbf{q}_2,\omega_2) \cdot\big[ \mathbf{\hat{v}}(\mathbf{k}-\mathbf{q})+\mathbf{\hat{v}}(\mathbf{k}-\mathbf{q}_1)\nonumber\\&&+\mathbf{\hat{v}}(\mathbf{k}+\mathbf{q}_2)+\mathbf{\hat{v}}(\mathbf{k})\big]\big\}|u_b(\mathbf{k})\rangle+\cdots\nonumber\\&\equiv&\langle u_a(\mathbf{k}-\mathbf{q}_1)|\mathbf{\hat{j}}(\mathbf{k}-\mathbf{q}_1,\mathbf{k})|u_b(\mathbf{k})\rangle,
\end{eqnarray}
\begin{eqnarray}\label{11}
\langle \psi_a(\mathbf{k}-\mathbf{q})|\mathbf{\hat{j}}(\mathbf{q})|\psi_b(\mathbf{k})\rangle&=&\langle u_a(\mathbf{k}-\mathbf{q})|-\frac{e}{2V} \big[ \mathbf{\hat{v}}(\mathbf{k}-\mathbf{q})+\mathbf{\hat{v}}(\mathbf{k})\big]|u_b(\mathbf{k})\rangle\nonumber\\&\equiv&\langle u_a(\mathbf{k}-\mathbf{q})|\mathbf{\hat{j}}(\mathbf{k}-\mathbf{q},\mathbf{k})|u_b(\mathbf{k})\rangle,
\end{eqnarray}
and 
\begin{eqnarray}\label{12}
\langle \psi_a(\mathbf{k})|\mathbf{\hat{j}}(\mathbf{q})|\psi_b(\mathbf{k})\rangle&=&\langle u_a(\mathbf{k})|-\frac{e}{2V} \big[ \partial_{\mathbf{k}}H^{'}(\mathbf{k},\mathbf{k}-\mathbf{q})+\partial_{\mathbf{k}}H^{'}(\mathbf{k}+\mathbf{q},\mathbf{k})\big]|u_b(\mathbf{k})\rangle\nonumber\\&=&-\frac{e}{2V}\langle u_a(\mathbf{k})|\partial_{\mathbf{k}}\bigg\{\frac{e}{2} \mathbf{A}(\mathbf{q},\omega) \cdot\big[ \mathbf{\hat{v}}(\mathbf{k}-\mathbf{q})+\mathbf{\hat{v}}(\mathbf{k})+(\mathbf{k}\leftrightarrow \mathbf{k}+\mathbf{q})\big]\nonumber\\&&+\bigg[\frac{e^2}{8} \mathbf{A}(\mathbf{q}_1,\omega_1)\cdot\partial_{\mathbf{p}}\big\{\mathbf{A}(\mathbf{q}_2,\omega_2) \cdot \big[\mathbf{\hat{v}}(\mathbf{k})+\mathbf{\hat{v}}(\mathbf{k}-\mathbf{q}_1)+\mathbf{\hat{v}}(\mathbf{k}-\mathbf{q}_2)\nonumber\\&&+\mathbf{\hat{v}}(\mathbf{k}-\mathbf{q})+(\mathbf{k}\leftrightarrow \mathbf{k}+\mathbf{q})\big]\big\}+(1\leftrightarrow 
2)\bigg]\bigg\}|u_b(\mathbf{k})\rangle+\cdots\nonumber\\
&\equiv&\langle u_a(\mathbf{k})|\mathbf{\hat{j}}(\mathbf{k},\mathbf{k})|u_b(\mathbf{k})\rangle,
\end{eqnarray}
where $H^{'}(\mathbf{k}+\mathbf{q},\mathbf{k})$ is given by replacing $\mathbf{k}$ in 
$H^{'}(\mathbf{k},\mathbf{k}-\mathbf{q})$ with $\mathbf{k}+\mathbf{q}$.

\section{Linear response}
\begin{figure}
\centering
\includegraphics[width=0.7\textwidth]{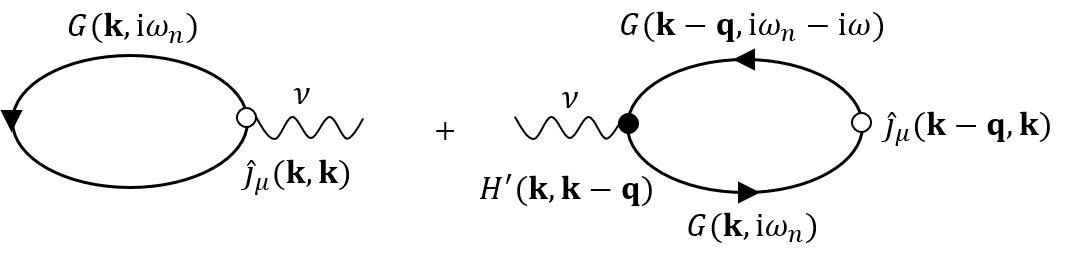}
\caption{The diagrams of linear response to vector potential. The wavy line $\nu$ in the hollow (or solid)  vertex indicates the presence of  a vector potential $A_{\nu}(\mathbf{q},\omega)$ in the current operator (or the interaction Hamiltonian). The second diagram illustrates the process of single-photon absorption.
}\label{figure.1}
\end{figure}

In this section, we  investigate the linear response with respect to the vector potential.
Based on the expression for the average current density, i.e.,  Eq.~(\ref{04}), and the matrix elements of the interaction Hamiltonian and the current operator, i.e.,  Eq.~(\ref{08}), Eq.~(\ref{11}), and  Eq.~(\ref{12}), we identify two terms that contribute to the linear response
\begin{eqnarray}\label{13}
\langle \hat{j}_{\mu}(\mathbf{q},i\omega) \rangle&=&\frac{1}{\beta }\sum_{\mathbf{k},i\omega_n}\mathrm{Tr}\bigg[\hat{j}_{\mu}(\mathbf{k},\mathbf{k})G(\mathbf{k},i\omega_n)+\hat{j}_{\mu}(\mathbf{k}-\mathbf{q},\mathbf{k})G(\mathbf{k},i\omega_n)H^{'}(\mathbf{k},\mathbf{k}-\mathbf{q})G(\mathbf{k}-\mathbf{q},i\omega_n-i\omega)\bigg]\nonumber\\&=&
\frac{-e^2}{4\beta V}\sum_{\mathbf{k},i\omega_n}\mathrm{Tr}\bigg[\partial_{\mu} \big[ 2\hat{v}_{\nu}(\mathbf{k})+\hat{v}_{\nu}(\mathbf{k}-\mathbf{q})+\hat{v}_{\nu}(\mathbf{k}+\mathbf{q})\big]G(\mathbf{k},i\omega_n)\bigg]A_{\nu}(\mathbf{q},\omega)\nonumber\\&&+\frac{-e^2}{4\beta V}\sum_{\mathbf{k},i\omega_n}\mathrm{Tr}\bigg[ \big[\hat{v}_{\mu}(\mathbf{k})+\hat{v}_{\mu}(\mathbf{k}-\mathbf{q})\big]G(\mathbf{k},i\omega_n)\big[\hat{v}_{\nu}(\mathbf{k})+\hat{v}_{\nu}(\mathbf{k}-\mathbf{q})\big]\nonumber\\&&\times G(\mathbf{k}-\mathbf{q},i\omega_n-i\omega)\bigg]A_{\nu}(\mathbf{q},\omega).
\end{eqnarray}
Here $\partial_{\mu}=\partial_{k_\mu}$, 
$G(\mathbf{k},i\omega_n)=\sum_a |u_a(\mathbf{k})\rangle\langle u_a(\mathbf{k})|/(i\omega_n+\mu-\varepsilon_a(\mathbf{k}))$, and  the repeated index $\nu$ is summed.
 Eq.~(\ref{13}) can be represented  by the  Feynman diagram shown  in Fig.~\ref{figure.1}. In the diagrams,  the solid line represents  the electron propagator, i.e., the  unperturbed Green's function, the wavy line refers to the vector potential. The  solid vertex denotes  the interaction Hamiltonian, while  the hollow vertex represents  the current operator.
 
Consider the current generated by the electric field $\mathbf{E}(\mathbf{q},\omega)=i\omega \mathbf{A}(\mathbf{q},\omega)$. Let $\mathbf{q}$ in Eq.~(\ref{13}) be zero, i.e., in spatially uniform field case, one has the conductivity
\begin{eqnarray}\label{14}
\sigma_{\mu\nu}(\omega) &=&\frac{ie^2}{\beta \omega V}\sum_{\mathbf{k},\omega_n}\mathrm{Tr}\big[ \partial_{\mu}\hat{v}_{\nu}(\mathbf{k})G(\mathbf{k},\omega_n)\big]+\frac{ie^2}{\beta \omega V}\sum_{\mathbf{k},\omega_n}\mathrm{Tr}\big[\hat{v}_{\mu}(\mathbf{k})G(\mathbf{k},\omega_n)\hat{v}_{\nu}(\mathbf{k}) G(\mathbf{k},\omega_n-\omega)\big].\nonumber\\
\end{eqnarray}
Here, $\omega$ represents the frequency of the incident photon, see Fig.~\ref{figure.1}. Upon performing the substitution $\omega \rightarrow -\omega$, we transition into a scenario involving an outgoing photon,  and can thereby obtain the corresponding conductivity.

Next, we consider the response to the magnetic field 
$\mathbf{B}(\mathbf{q},\omega)=i\mathbf{q}\times \mathbf{A}(\mathbf{q},\omega)$. 
To analyze this response, we need to expand the coefficients in Eq.~(\ref{13})
to first order in $\mathbf{q}$ ~\cite{araki2021spin}. Note that the coefficient in the first term  is an even function of the wave vector $\mathbf{q}$. As a result, this term  does not contribution to such a response. The Matsubara summation in Eq.~(\ref{13}) can be evaluated as  follows
\begin{eqnarray}\label{141}
\frac{1}{\beta}\sum_{i\omega_n}\frac{1}{i\omega_n-\varepsilon_a(\mathbf{k})}\frac{1}{i\omega_n-i\omega-\varepsilon_b(\mathbf{k}-\mathbf{q})}=\frac{f(\varepsilon_a(\mathbf{k}))-f(\varepsilon_b(\mathbf{k}-\mathbf{q}))}{\varepsilon_a(\mathbf{k})-\varepsilon_b(\mathbf{k}-\mathbf{q})-i\omega}\equiv F_{ab}(\mathbf{k},\mathbf{q},i\omega),
\end{eqnarray}
where $f(\varepsilon_a(\mathbf{k}))=1/\big(e^{\beta(\varepsilon_a(\mathbf{k})-\mu)}+1\big)$ is the Fermi-Dirac distribution function.
After taking the analytical continuation $i\omega\rightarrow \omega+i 0$, we can rewrite 
the response as
\begin{eqnarray}\label{15}
\langle \hat{j}_{\mu}(\mathbf{q},\omega) \rangle=\Pi_{\mu\nu}(\mathbf{q},\omega)A_{\nu}(\mathbf{q},\omega),
\end{eqnarray}
where 
\begin{eqnarray}\label{16}
\Pi_{\mu\nu}(\mathbf{q},\omega)=\frac{-e^2}{4 V}\sum_{\mathbf{k},a,b}F_{ab}(\mathbf{k},\mathbf{q},\omega)M_{ab}(\mathbf{k},\mathbf{q})
\end{eqnarray}
and
\begin{eqnarray}\label{17}
M_{ab}(\mathbf{k},\mathbf{q})=\langle u_b(\mathbf{k}-\mathbf{q})|\hat{v}_{\mu}(\mathbf{k})+\hat{v}_{\mu}(\mathbf{k}-\mathbf{q})|u_a(\mathbf{k})\rangle\langle u_a(\mathbf{k})|\hat{v}_{\nu}(\mathbf{k})+\hat{v}_{\nu}(\mathbf{k}-\mathbf{q})|u_b(\mathbf{k}-\mathbf{q})\rangle.
\end{eqnarray}
It can be easily found that the response function $\Pi_{\mu\nu}(\mathbf{q},\omega)$ is Hermitean, i.e.,
\begin{eqnarray}\label{18}
\Pi_{\mu\nu}(\mathbf{q},\omega)=\Pi^{*}_{\nu\mu}(\mathbf{q},\omega).
\end{eqnarray}
Moreover, this function satisfies the following relation
\begin{eqnarray}\label{19}
\Pi_{\mu\nu}(\mathbf{q},\omega)=\Pi^{*}_{\mu\nu}(-\mathbf{q},-\omega),
\end{eqnarray}
which can be proved by the interchanging  $a \leftrightarrow b$ and replacing 
$\mathbf{k} \rightarrow \mathbf{k}-\mathbf{q}$ in $\Pi^{*}_{\mu\nu}(-\mathbf{q},-\omega)$.
By expanding Eq.~(\ref{18}) and Eq.~(\ref{19})  in terms of $\mathbf{q}$ and  comparing the coefficients at the first order, we obtain
\begin{eqnarray}\label{20}
\partial_{q_{\eta}}\Pi_{\mu\nu}(\mathbf{q},\omega)|_{q_{\eta}=0}=\partial_{q_{\eta}}\Pi^{*}_{\nu\mu}(\mathbf{q},\omega)|_{q_{\eta}=0}
=-\partial_{q_{\eta}}\Pi^{*}_{\mu\nu}(\mathbf{q},-\omega)|_{q_{\eta}=0}.
\end{eqnarray}
Here, the wave vector $\mathbf{q}$ is selected to be oriented along the $\eta$ axis.
Now we have a response
\begin{eqnarray}\label{15t}
\langle \hat{j}_{\mu}(\mathbf{q},\omega) \rangle=\partial_{q_{\eta}}\Pi_{\mu\nu}(\mathbf{q},\omega)|_{q_{\eta}=0}q_{\eta}A_{\nu}(\mathbf{q},\omega)\equiv \alpha_{\mu\rho}(\omega)B_{\rho}(\mathbf{q},\omega),
\end{eqnarray}
where \begin{eqnarray}\label{15c}
\alpha_{\mu\rho}(\omega)=i\epsilon_{\rho\nu\eta}\partial_{q_{\eta}}\Pi_{\mu\nu}(\mathbf{q},\omega)|_{q_{\eta}=0}
\end{eqnarray}
is the response coefficient and 
$B_{\rho}=i\epsilon_{\rho\eta\nu}q_{\eta}A_{\nu}(\mathbf{q},\omega)$ is the magnetic field. 
In the limit $\omega\rightarrow 0$,  one can deduce from Eq.~(\ref{20}) that   the function  $\partial_{q_{\eta}}\Pi_{\mu\nu}(\mathbf{q},0)|_{q_{\eta}=0}$ satisfies the following relation
\begin{eqnarray}\label{23}
\partial_{q_{\eta}}\Pi_{\mu\nu}(\mathbf{q},0)|_{q_{\eta}=0}=i\mathrm{Im}\partial_{q_{\eta}}\Pi_{\mu\nu}(\mathbf{q},0)|_{q_{\eta}=0}=-i\mathrm{Im}\partial_{q_{\eta}}\Pi_{\nu\mu}(\mathbf{q},0)|_{q_{\eta}=0}.
\end{eqnarray}
In this paper, we use the uniform limit, i.e., setting $\mathbf{q}\rightarrow 0$ before $\omega\rightarrow 0$, since the static limit, i.e., sending $\omega\rightarrow 0$ before $\mathbf{q}\rightarrow 0$, often yields a null result ~\cite{mahan2013many,chang2015chiral}.
The function $\partial_{q_{\eta}}\Pi_{\mu\nu}(\mathbf{q},0)|_{q_{\eta}=0}$ can be decomposed into intraband and interband parts. Due to the fact that 
$F_{aa}(\mathbf{k},0,\omega)$ and the imaginary part of $M_{aa}(\mathbf{k},0)$ are both  zero,
the intraband part, i.e., $-ie^2\mathrm{lim}_{\omega\rightarrow 0}\sum_{\mathbf{k},a}\big[\partial_{\eta}F_{aa}(\mathbf{k},\mathbf{q},\omega)|_{q_{\eta}=0}\mathrm{Im}M_{aa}(\mathbf{k},0)+F_{aa}(\mathbf{k},0,\omega)\partial_{\eta}\mathrm{Im}M_{aa}(\mathbf{k},\mathbf{q})|_{q_{\eta}=0}\big]/4 V$,
vanishes in the uniform limit \cite{zhong2016gyrotropic}. By replacing the wave vector $\mathbf{k}$ in Eq.~(\ref{16}) with  $\mathbf{k}+ \mathbf{q}/2$ and exchanging the indices $a\leftrightarrow b$ in terms containing $f(\varepsilon_b(\mathbf{k}))$, we are able to derive the interband part of $\partial_{q_{\eta}}\Pi_{\mu\nu}(\mathbf{q},0)|_{q_{\eta}=0}$, which subsequently allows us to determine the response coefficient
\begin{eqnarray}\label{24}
\alpha_{\mu\rho}(0)&=&\frac{-e^2\epsilon_{\rho\nu\eta}}{2V}\sum_{\mathbf{k},a}\big\{f(\varepsilon_a(\mathbf{k}))\big[F^{ab}_{\mu\nu}(v_{a\eta}+v_{b\eta})-(\eta\leftrightarrow\mu)-(\eta\leftrightarrow\nu)\big]-\partial_{\eta}f(\varepsilon_a(\mathbf{k}))F^{ab}_{\mu\nu}\varepsilon_{ab}\big\}\nonumber\\&=&\frac{-e^2}{2V}\sum_{\mathbf{k},a}\big[2f(\varepsilon_a(\mathbf{k})) F^{ab}_{\lambda}(v_{a\lambda}+v_{b\lambda})\delta_{\mu\rho}-\epsilon_{\rho\nu\eta}\partial_{\eta}f(\varepsilon_a(\mathbf{k}))F^{ab}_{\mu\nu}\varepsilon_{ab}\big],
\end{eqnarray}
where $F^{ab}_{\mu\nu}=-2\mathrm{lm}\langle \partial_{\mu}u_a(\mathbf{k})\vert  u_b(\mathbf{k})\rangle \langle u_b(\mathbf{k})\vert  \partial_{\nu} u_a(\mathbf{k})\rangle$ is the Berry curvature,  $F^{ab}_{\lambda}=\epsilon_{\mu\nu\lambda}F^{ab}_{\mu\nu}/2$, $v_{a\lambda}=\partial_{\lambda}\varepsilon_a(\mathbf{k})$ is the group velocity along $\lambda$-axis, and  $\varepsilon_{ab}\equiv\varepsilon_{a}(\mathbf{k})-\varepsilon_{b}(\mathbf{k})$.
Here, the relation $\langle u_b(\mathbf{k})\vert \hat{v}_{\mu}(\mathbf{k})\vert  u_a(\mathbf{k})\rangle=\varepsilon_{ab}\langle u_b(\mathbf{k})\vert \partial_{\mu}  u_a(\mathbf{k})\rangle+v_{a\mu}\delta_{ab}$
is employed and the Einstein summation convention is applied to  the index $\lambda$.

\section{formalism of Nonlinear response}
In the following, we investigate the nonlinear response with respect to the vector potential, focusing specifically on the second-order response. By combining Eq.~(\ref{04}) and Eqs.~(\ref{08})-(\ref{12}), 
we identify  four terms that contribute to the current $\langle \hat{j}_{\mu}(\mathbf{q},i\omega) \rangle$, associated with $A_{\nu}(\mathbf{q}_1,\omega_1)A_{\gamma}(\mathbf{q}_2,\omega_2)$
\begin{eqnarray}\label{25}
\langle \hat{j}^{(1)}_{\mu}(\mathbf{q},i\omega) \rangle&=&\frac{1}{\beta}\sum_{\mathbf{k},i\omega_n}\mathrm{Tr}\bigg[\hat{j}_{\mu}(\mathbf{k},\mathbf{k})G(\mathbf{k},i\omega_n)\bigg]\nonumber\\
&=&\frac{-e^3}{16\beta V}\sum_{\mathbf{k},i\omega_n}\mathrm{Tr} \bigg[ \partial_{\mu}\partial_{\nu}\big[\hat{v}_{\gamma}(\mathbf{k})+\hat{v}_{\gamma}(\mathbf{k}-\mathbf{q}_1)+\hat{v}_{\gamma}(\mathbf{k}-\mathbf{q}_2)+\hat{v}_{\gamma}(\mathbf{k}-\mathbf{q})+(\mathbf{k}\leftrightarrow \mathbf{k}+\mathbf{q})\big]\nonumber\\&&\times G(\mathbf{k},i\omega_n)+(1,\nu\leftrightarrow 2,\gamma)\bigg]A_{\nu}(\mathbf{q}_1,\omega_1)A_{\gamma}(\mathbf{q}_2,\omega_2),
\end{eqnarray}
\begin{eqnarray}\label{26}
\langle \hat{j}^{(2)}_{\mu}(\mathbf{q},i\omega) \rangle&=&\frac{1}{\beta}\sum_{\mathbf{k},i\omega_n}\mathrm{Tr}\bigg[\hat{j}_{\mu}(\mathbf{k}-\mathbf{q}_1,\mathbf{k})G(\mathbf{k},i\omega_n)H^{'}(\mathbf{k},\mathbf{k}-\mathbf{q}_1)G(\mathbf{k}-\mathbf{q}_1,i\omega_n-i\omega_1)\bigg]+(1\leftrightarrow 2)\nonumber\\
&=&\frac{-e^3}{8\beta V}\sum_{\mathbf{k},i\omega_n}\mathrm{Tr} \bigg[ \partial_{\mu}\big[\hat{v}_{\gamma}(\mathbf{k})+\hat{v}_{\gamma}(\mathbf{k}-\mathbf{q}_1)+\hat{v}_{\gamma}(\mathbf{k}+\mathbf{q}_2)+\hat{v}_{\gamma}(\mathbf{k}-\mathbf{q})\big]G(\mathbf{k},i\omega_n)\nonumber\\&&\times\big[\hat{v}_{\nu}(\mathbf{k})+\hat{v}_{\nu}(\mathbf{k}-\mathbf{q}_1)\big] G(\mathbf{k}-\mathbf{q}_1,i\omega_n-i\omega_1)+(1,\nu\leftrightarrow 2,\gamma)\bigg]A_{\nu}(\mathbf{q}_1,\omega_1)A_{\gamma}(\mathbf{q}_2,\omega_2),\nonumber\\
\end{eqnarray}
\begin{eqnarray}\label{27}
\langle \hat{j}^{(3)}_{\mu}(\mathbf{q},i\omega) \rangle&=&\frac{1}{\beta}\sum_{\mathbf{k},i\omega_n}\mathrm{Tr}\bigg[\hat{j}_{\mu}(\mathbf{k}-\mathbf{q},\mathbf{k})G(\mathbf{k},i\omega_n)H^{'}(\mathbf{k},\mathbf{k}-\mathbf{q})G(\mathbf{k}-\mathbf{q},i\omega_n-i\omega)\bigg]\nonumber\\
&=&\frac{-e^3}{16\beta V}\sum_{\mathbf{k},i\omega_n}\mathrm{Tr} \bigg[ \big[\hat{v}_{\mu}(\mathbf{k})+\hat{v}_{\mu}(\mathbf{k}-\mathbf{q})\big]G(\mathbf{k},i\omega_n)\partial_{\nu}\big[\hat{v}_{\gamma}(\mathbf{k})+\hat{v}_{\gamma}(\mathbf{k}-\mathbf{q}_1)\nonumber\\&&+\hat{v}_{\gamma}(\mathbf{k}-\mathbf{q}_2)+\hat{v}_{\gamma}(\mathbf{k}-\mathbf{q})\big]  G(\mathbf{k}-\mathbf{q},i\omega_n-i\omega)+(1,\nu\leftrightarrow 2,\gamma)\bigg]A_{\nu}(\mathbf{q}_1,\omega_1)A_{\gamma}(\mathbf{q}_2,\omega_2),\nonumber\\
\end{eqnarray}
and 
\begin{eqnarray}\label{28}
\langle \hat{j}^{(4)}_{\mu}(\mathbf{q},i\omega) \rangle&=&\frac{1}{\beta}\sum_{\mathbf{k},i\omega_n}\mathrm{Tr}\bigg[\hat{j}_{\mu}(\mathbf{k}-\mathbf{q},\mathbf{k})G(\mathbf{k},i\omega_n)H^{'}(\mathbf{k},\mathbf{k}-\mathbf{q}_1)G(\mathbf{k}-\mathbf{q}_1,i\omega_n-i\omega_1)\nonumber\\&& \times H^{'}(\mathbf{k}-\mathbf{q}_1,\mathbf{k}-\mathbf{q})G(\mathbf{k}-\mathbf{q},i\omega_n-i\omega)\bigg]+(1\leftrightarrow 2)\nonumber\\
&=&\frac{-e^3}{8\beta V}\sum_{\mathbf{k},i\omega_n}\mathrm{Tr} \bigg[ \big[\hat{v}_{\mu}(\mathbf{k})+\hat{v}_{\mu}(\mathbf{k}-\mathbf{q})\big]G(\mathbf{k},i\omega_n)\big[\hat{v}_{\nu}(\mathbf{k})+\hat{v}_{\nu}(\mathbf{k}-\mathbf{q}_1)\big] G(\mathbf{k}-\mathbf{q}_1,i\omega_n-i\omega_1)\nonumber\\&&\times \big[\hat{v}_{\gamma}(\mathbf{k}-\mathbf{q}_1)+\hat{v}_{\gamma}(\mathbf{k}-\mathbf{q})\big]G(\mathbf{k}-\mathbf{q},i\omega_n-i\omega)+(1,\nu\leftrightarrow 2,\gamma)\bigg]A_{\nu}(\mathbf{q}_1,\omega_1)A_{\gamma}(\mathbf{q}_2,\omega_2).\nonumber\\
\end{eqnarray}
\begin{figure}
\centering
\includegraphics[width=0.9\textwidth]{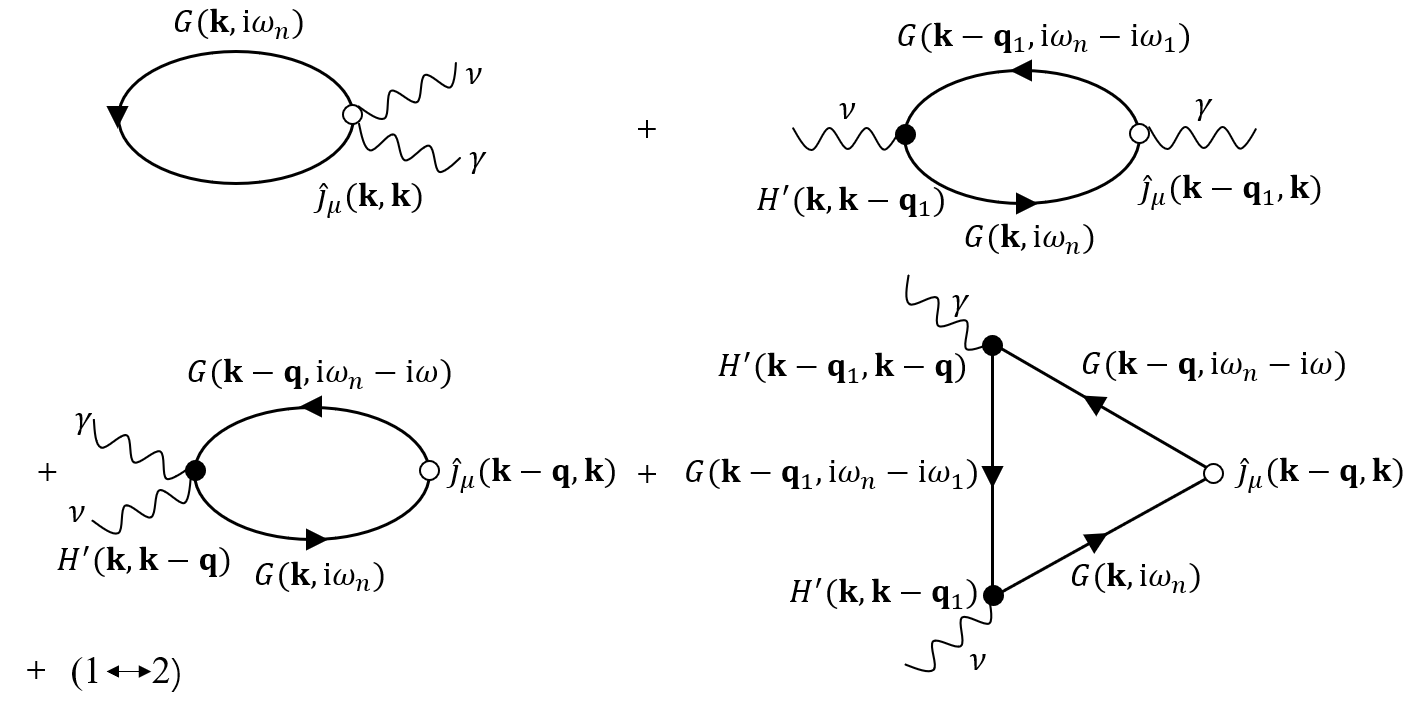}
\caption{The diagrams of second-order response to vector potential. The wavy lines $\nu$ and $\gamma$ denote the vector potentials $A_{\nu}(\mathbf{q}_1,\omega_1)$ and $A_{\gamma}(\mathbf{q}_2,\omega_2)$, respectively. The notation $(1\leftrightarrow 2)$ signifies the  interchange  of the indices $1$ and $2$ in the second and fourth diagrams. The second diagram portrays  the process of single-photon absorption, while  the third and fourth diagrams depict two-photon absorption.
}\label{figure.2}
\end{figure}
Here, the repeated indices $\nu,\gamma$ are implicitly summed over. This second-order response can be represented by the  Feynman diagram shown  in Fig.~\ref{figure.2}.

After  taking the Matsubara summation and analytical continuation, the  coefficients in the above equations can be  written respectively as follows
\begin{eqnarray}\label{3001z}
\Pi^{(1)}_{\mu\nu\gamma}(\mathbf{q}_1,\mathbf{q}_2,0,0)=\frac{-e^3}{16 V}\sum_{\mathbf{k},a,b}f(\varepsilon_a(\mathbf{k}))M^{(1)}_{ab}(\mathbf{k},\mathbf{q}_1,\mathbf{q}_2)+(1,\nu\leftrightarrow 2,\gamma),
\end{eqnarray}
\begin{eqnarray}\label{30}
\Pi^{(2)}_{\mu\nu\gamma}(\mathbf{q}_1,\mathbf{q}_2,\omega_1,\omega_2)=\frac{-e^3}{8 V}\sum_{\mathbf{k},a,b}F_{ab}(\mathbf{k},\mathbf{q}_1,\omega_1)M^{(2)}_{ab}(\mathbf{k},\mathbf{q}_1,\mathbf{q}_2)+(1,\nu\leftrightarrow 2,\gamma),
\end{eqnarray}
\begin{eqnarray}\label{31}
\Pi^{(3)}_{\mu\nu\gamma}(\mathbf{q}_1,\mathbf{q}_2,\omega_1,\omega_2)=\frac{-e^3}{16 V}\sum_{\mathbf{k},a,b}F_{ab}(\mathbf{k},\mathbf{q},\omega)M^{(3)}_{ab}(\mathbf{k},\mathbf{q}_1,\mathbf{q}_2)+(1,\nu\leftrightarrow 2,\gamma),
\end{eqnarray}
and 
\begin{eqnarray}\label{32}
\Pi^{(4)}_{\mu\nu\gamma}(\mathbf{q}_1,\mathbf{q}_2,\omega_1,\omega_2)=\frac{-e^3}{8 V}\sum_{\mathbf{k},a,b,c}F_{abc}(\mathbf{k},\mathbf{q},\omega)M^{(4)}_{abc}(\mathbf{k},\mathbf{q}_1,\mathbf{q}_2)+(1,\nu\leftrightarrow 2,\gamma),
\end{eqnarray}
where 
$F_{ab}(\mathbf{k},\mathbf{q}_1,\omega_1)$ and $F_{ab}(\mathbf{k},\mathbf{q},\omega)$ are defined in Eq.~(\ref{141}), 
\begin{eqnarray}\label{33}
F_{abc}(\mathbf{k},\mathbf{q},\omega)&=&\frac{1}{\varepsilon_a(\mathbf{k})-\varepsilon_b(\mathbf{k}-\mathbf{q}_1)-\omega_1}\bigg[\frac{f(\varepsilon_a(\mathbf{k}))-f(\varepsilon_c(\mathbf{k}-\mathbf{q}))}{\varepsilon_a(\mathbf{k})-\varepsilon_c(\mathbf{k}-\mathbf{q})-\omega}-\frac{f(\varepsilon_b(\mathbf{k}-\mathbf{q}_1))-f(\varepsilon_c(\mathbf{k}-\mathbf{q}))}{\varepsilon_b(\mathbf{k}-\mathbf{q}_1)-\varepsilon_c(\mathbf{k}-\mathbf{q})-\omega_2}\bigg],\nonumber\\
\end{eqnarray}
\begin{eqnarray}\label{3001}
M^{(1)}_{ab}(\mathbf{k},\mathbf{q}_1,\mathbf{q}_2)&=&
\langle u_a(\mathbf{k})|\partial_{\mu}\partial_{\nu}\big[\hat{v}_{\gamma}(\mathbf{k})+\hat{v}_{\gamma}(\mathbf{k}-\mathbf{q}_1)+\hat{v}_{\gamma}(\mathbf{k}-\mathbf{q}_2)+\hat{v}_{\gamma}(\mathbf{k}-\mathbf{q})+(\mathbf{k}\leftrightarrow \mathbf{k}+\mathbf{q})|u_a(\mathbf{k})\rangle,\nonumber\\
\end{eqnarray}
\begin{eqnarray}\label{301}
M^{(2)}_{ab}(\mathbf{k},\mathbf{q}_1,\mathbf{q}_2)&=&\langle u_b(\mathbf{k}-\mathbf{q}_1)|\partial_{\mu}\big[\hat{v}_{\gamma}(\mathbf{k})+\hat{v}_{\gamma}(\mathbf{k}-\mathbf{q}_1)+\hat{v}_{\gamma}(\mathbf{k}+\mathbf{q}_2)+\hat{v}_{\gamma}(\mathbf{k}-\mathbf{q})\big]|u_a(\mathbf{k})\rangle\nonumber\\&&\times\langle u_a(\mathbf{k})|\hat{v}_{\nu}(\mathbf{k})+\hat{v}_{\nu}(\mathbf{k}-\mathbf{q}_1)|u_b(\mathbf{k}-\mathbf{q}_1)\rangle,
\end{eqnarray}
\begin{eqnarray}\label{302}
M^{(3)}_{ab}(\mathbf{k},\mathbf{q}_1,\mathbf{q}_2)&=&\langle u_b(\mathbf{k}-\mathbf{q})|\hat{v}_{\mu}(\mathbf{k})+\hat{v}_{\mu}(\mathbf{k}-\mathbf{q})|u_a(\mathbf{k})\rangle\nonumber\\&&\times\langle u_a(\mathbf{k})|\partial_{\nu}\big[\hat{v}_{\gamma}(\mathbf{k})+\hat{v}_{\gamma}(\mathbf{k}-\mathbf{q}_1)+\hat{v}_{\gamma}(\mathbf{k}-\mathbf{q}_2)+\hat{v}_{\gamma}(\mathbf{k}-\mathbf{q})\big]|u_b(\mathbf{k}-\mathbf{q})\rangle,
\end{eqnarray}
and
\begin{eqnarray}\label{303}
M^{(4)}_{abc}(\mathbf{k},\mathbf{q}_1,\mathbf{q}_2)&=&\langle u_c(\mathbf{k}-\mathbf{q})|\hat{v}_{\mu}(\mathbf{k})+\hat{v}_{\mu}(\mathbf{k}-\mathbf{q})|u_a(\mathbf{k})\rangle\langle u_a(\mathbf{k})|\hat{v}_{\nu}(\mathbf{k})+\hat{v}_{\nu}(\mathbf{k}-\mathbf{q}_1)|u_b(\mathbf{k}-\mathbf{q}_1)\rangle\nonumber\\&&\times
\langle u_b(\mathbf{k}-\mathbf{q}_1)|\hat{v}_{\gamma}(\mathbf{k}-\mathbf{q}_1)+\hat{v}_{\gamma}(\mathbf{k}-\mathbf{q}))|u_c(\mathbf{k}-\mathbf{q})\rangle.
\end{eqnarray}
By exchanging the indices $a\leftrightarrow b$ in 
$\Pi^{(2,3)*}_{\mu\nu\gamma}(-\mathbf{q}_1,-\mathbf{q}_2,-\omega_1,-\omega_2)$, $a\leftrightarrow c$ in 
$\Pi^{(4)*}_{\mu\nu\gamma}(-\mathbf{q}_1,-\mathbf{q}_2,-\omega_1,-\omega_2)$, and taking the replacement $\mathbf{k}\rightarrow \mathbf{k}-\mathbf{q}_1$ in 
$\Pi^{(2)*}_{\mu\nu\gamma}(-\mathbf{q}_1,-\mathbf{q}_2,-\omega_1,-\omega_2)$, $\mathbf{k}\rightarrow \mathbf{k}-\mathbf{q}$ in 
$\Pi^{(3,4)*}_{\mu\nu\gamma}(-\mathbf{q}_1,-\mathbf{q}_2,-\omega_1,-\omega_2)$, one can derive  the relation
\begin{eqnarray}\label{34}
\Pi^{(i)}_{\mu\nu\gamma}(\mathbf{q}_1,\mathbf{q}_2,\omega_1,\omega_2)=\Pi^{(i)*}_{\mu\nu\gamma}(-\mathbf{q}_1,-\mathbf{q}_2,-\omega_1,-\omega_2),
\end{eqnarray}
which  represents a   general constraint governing transport coefficients.

\section{Nonlinear Hall effects}
In the subsequent analysis, we employ the theoretical framework previously established to derive several key results. Specifically, we examine the nonlinear Hall effects in both time-reversal symmetric and time-reversal breaking systems.  Our analysis encompasses both the general case and the specific scenario of
SHG. 

Consider the second-order response with respect to 
$E_{\nu}(\mathbf{q}_1,\omega_1)E_{\gamma}(\mathbf{q}_2,\omega_2)=-\omega_1\omega_2A_{\nu}(\mathbf{q}_1,\omega_1)A_{\gamma}(\mathbf{q}_2,\omega_2)$. In the uniform field case, i.e., $\mathbf{q}_i=0$, it follows from  Eqs.~(\ref{25})-(\ref{28})
that the second-order conductivity is given by
\begin{eqnarray}\label{29}
\sigma_{\mu\nu\gamma
}(\omega_1,\omega_2)&=&\frac{e^3}{\omega_1\omega_2\beta V}\sum_{\mathbf{k},\omega_n}\mathrm{Tr} \bigg[ \frac{1}{2}\partial_{\mu}\partial_{\nu}\hat{v}_{\gamma}(\mathbf{k})G(\mathbf{k},\omega_n)+\partial_{\mu}\hat{v}_{\gamma}(\mathbf{k})G(\mathbf{k},\omega_n)\hat{v}_{\nu}(\mathbf{k})G(\mathbf{k},\omega_n-\omega_1)\nonumber\\&&+\frac{1}{2}\hat{v}_{\mu}(\mathbf{k})G(\mathbf{k},\omega_n)\partial_{\nu}\hat{v}_{\gamma}(\mathbf{k})G(\mathbf{k},\omega_n-\omega)\nonumber\\&&+\hat{v}_{\mu}G(\mathbf{k},\omega_n)\hat{v}_{\nu}G(\mathbf{k},\omega_n-\omega_1)\hat{v}_{\gamma}G(\mathbf{k},\omega_n-\omega)\bigg]+(1,\nu\leftrightarrow 2,\gamma),
\end{eqnarray}
where $\omega_1$ and $\omega_2$ represent the frequencies of the incident photons. By performing the substitution $\omega_i \rightarrow -\omega_i$, we get the conductivity in a scenario characterized by the presence of  outgoing photons~\cite{parker2019diagrammatic}. It should be noted that we use -e  rather than e~\cite{parker2019diagrammatic} to represent the electron charge. 

From  Eqs.~(\ref{25}) to (\ref{32}), it can be inferred that the aforementioned  conductivity, i.e., Eqs.~(\ref{29}), can also be expressed as
\begin{eqnarray}\label{qt}
\sigma_{\mu\nu\gamma
}(\omega_1,\omega_2)=\frac{-1}{\omega_1\omega_2}\sum^4_{i=1}\Pi^{(i)}_{\mu\nu\gamma}(0,0,\omega_1,\omega_2).
\end{eqnarray}
By  performing a Taylor expansion of $\Pi^{(i)}_{\mu\nu\gamma}(0,0,\omega_1,\omega_2)$  with respect to $\omega_1$ and $\omega_2$, we have
\begin{eqnarray}\label{qq}
\Pi^{(i)}_{\mu\nu\gamma}(0,0,\omega_1,\omega_2)&=&\Pi^{(i)}_{\mu\nu\gamma}(0,0,0,0)+\partial_{\omega_1}\Pi^{(i)}_{\mu\nu\gamma}(0,0,\omega_1,0)|_{\omega_1=0}\omega_1+\partial_{\omega_2}\Pi^{(i)}_{\mu\nu\gamma}(0,0,0,\omega_2)|_{\omega_2=0}\omega_2\nonumber\\&&+\partial_{\omega_2}\partial_{\omega_1}\Pi^{(i)}_{\mu\nu\gamma}(0,0,\omega_1,\omega_2)|_{\omega_1,\omega_2=0}\omega_1\omega_2+\cdots.
\end{eqnarray}
Upon substituting Eq.~(\ref{qq}) into Eq.~(\ref{qt}), it becomes evident that three distinct types of conductivity emerge. Specifically, the zeroth-order term of Eq.~(\ref{qq}) results in a contribution containing $1/\omega_1\omega_2$, which is not of particular interest to us. The second (third) term gives rise to a contribution denoted by $\sigma^{(1,0)}_{\mu\nu\gamma
}$ $\big(\sigma^{(0,1)}_{\mu\nu\gamma
}\big)$. The conductivity $\sigma^{(1,0)}_{\mu\nu\gamma
}+\sigma^{(0,1)}_{\mu\nu\gamma
}$
is frequency-dependent and  can be regarded as an extrinsic response.  Lastly, the final term yields a contribution represented by $\sigma^{(1,1)}_{\mu\nu\gamma
}$, which is independent of frequency and thus represents an intrinsic response. Here, the superscripts in the conductivities respectively indicate  the orders of $\omega_1$ and $\omega_2$ in  Eq.~(\ref{qq}).
By expanding the relation  in Eq.~(\ref{34}) and comparing the coefficients at the same order, we can derive 
\begin{eqnarray}\label{43}
\partial_{\omega_1}\Pi^{(i)}_{\mu\nu\gamma}(0,0,\omega_1,0)|_{\omega_1=0}&=&-\partial_{\omega_1}\Pi^{(i)*}_{\mu\nu\gamma}(0,0,\omega_1,0)|_{\omega_1=0}\nonumber\\&=&i\mathrm{Im}\partial_{\omega_1}\Pi^{(i)}_{\mu\nu\gamma}(0,0,\omega_1,0)|_{\omega_1=0},
\end{eqnarray}
and
\begin{eqnarray}\label{43t}
\partial_{\omega_2}\partial_{\omega_1}\Pi^{(i)}_{\mu\nu\gamma}(0,0,\omega_1,\omega_2)|_{\omega_1,\omega_2=0}&=&\partial_{\omega_2}\partial_{\omega_1}\Pi^{(i)*}_{\mu\nu\gamma}(0,0,\omega_1,\omega_2)|_{\omega_1,\omega_2=0}\nonumber\\&=&\mathrm{Re}\partial_{\omega_2}\partial_{\omega_1}\Pi^{(i)}_{\mu\nu\gamma}(0,0,\omega_1,\omega_2)|_{\omega_1,\omega_2=0}.
\end{eqnarray}
Consequently, the conductivity $\sigma^{(1,0)}_{\mu\nu\gamma
}$ and $\sigma^{(1,1)}_{\mu\nu\gamma
}$ can be respectively formulated  as
\begin{eqnarray}\label{44}
\sigma^{(1,0)}_{\mu\nu\gamma
}&=&\frac{1}{i\omega_2}\sum^4_{i=2}\mathrm{Im}\partial_{\omega_1}\Pi^{(i)}_{\mu\nu\gamma}(0,0,\omega_1,0)|_{\omega_1=0},
\end{eqnarray}
and
\begin{eqnarray}\label{45}
\sigma^{(1,1)}_{\mu\nu\gamma
}&=&-\sum^4_{i=2}\mathrm{Re}\partial_{\omega_2}\partial_{\omega_1}\Pi^{(i)}_{\mu\nu\gamma}(0,0,\omega_1,\omega_2)|_{\omega_1,\omega_2=0}.
\end{eqnarray}
By employing a similar procedure, $\sigma^{(0,1)}_{\mu\nu\gamma
}$ can also be obtained.
Note that in the above expression, the term for $i=1$ is independent of frequency, as can be seen from Eq.~(\ref{3001z}), so this term does not contribute to these two responses.
The above analysis applies to situations where $\omega_1$ and $\omega_2$ are not related. 
When  $\omega_1=\omega_2=\omega_0$, i.e., in the case of SHG, Eq.~(\ref{34}) becomes 
\begin{eqnarray}\label{340}
\Pi^{(i)}_{\mu\nu\gamma}(0,0,\omega_0,\omega_0)=\Pi^{(i)*}_{\mu\nu\gamma}(0,0,-\omega_0,-\omega_0).
\end{eqnarray}
By performing a Taylor expansion on this relation up to second order of $\omega_0$ and comparing the coefficients, we can have
\begin{eqnarray}\label{qqt}
\Pi^{(i)}_{\mu\nu\gamma}(0,0,\omega_0,\omega_0)&=&\Pi^{(i)}_{\mu\nu\gamma}(0,0,0,0)+\partial_{\omega_0}\Pi^{(i)}_{\mu\nu\gamma}(0,0,\omega_0,\omega_0)|_{\omega_0=0}\omega_0\nonumber\\&+&\frac{1}{2}\partial^2_{\omega_0}\Pi^{(i)}_{\mu\nu\gamma}(0,0,\omega_0,\omega_0)|_{\omega_0=0}\omega^2_0,
\end{eqnarray}
\begin{eqnarray}
\partial_{\omega_0}\Pi^{(i)}_{\mu\nu\gamma}(0,0,\omega_0,\omega_0)|_{\omega_0=0}&=&i\mathrm{Im}\partial_{\omega_0}\Pi^{(i)}_{\mu\nu\gamma}(0,0,\omega_0,\omega_0)|_{\omega_0=0},
\end{eqnarray}
and
\begin{eqnarray}
\partial^2_{\omega_0}\Pi^{(i)}_{\mu\nu\gamma}(0,0,\omega_0,\omega_0)|_{\omega_0=0}&=&\mathrm{Re}\partial^2_{\omega_0}\Pi^{(i)}_{\mu\nu\gamma}(0,0,\omega_0,\omega_0)|_{\omega_0=0}.
\end{eqnarray}
As a consequence, the aforementioned two types of conductivity are correspondingly revised to
\begin{eqnarray}\label{44a}
\sigma^{(1)}_{\mu\nu\gamma}&=&\frac{1}{i\omega_0}\sum^4_{i=2}\mathrm{Im}\partial_{\omega_0}\Pi^{(i)}_{\mu\nu\gamma}(0,0,\omega_0,\omega_0)|_{\omega_0=0},
\end{eqnarray}
and
\begin{eqnarray}\label{45a}
\sigma^{(2)}_{\mu\nu\gamma}&=&\frac{-1}{2}\sum^4_{i=2}\mathrm{Re}\partial^2_{\omega_0}\Pi^{(i)}_{\mu\nu\gamma}(0,0,\omega_0,\omega_0)|_{\omega_0=0}.
\end{eqnarray}
Here, the superscripts in the conductivities, namely  Eqs.~(\ref{44a}) and (\ref{45a}), respectively denote   the orders of $\omega_0$  in  Eq.~(\ref{qqt}).
In general, conductivity will give different values as the frequencies of the fields approach  zero along different ways, i.e., $\sigma^{(1,0)}_{\mu\nu\gamma}+\sigma^{(0,1)}_{\mu\nu\gamma}\neq \sigma^{(1)}_{\mu\nu\gamma}$ and $\sigma^{(1,1)}_{\mu\nu\gamma}\neq \sigma^{(2)}_{\mu\nu\gamma}$. It is worth noting that
since we use Taylor expansion to derive different types of conductivities, these conductivities naturally share the same dimension.

\subsection{Nonlinear Hall effect in time-reversal symmetric systems}
Let us consider the response in  Eq.~(\ref{44}). From  the previous  section, we have
\begin{eqnarray}\label{46}
\sum^4_{i=2}\Pi^{(i)}_{\mu\nu\gamma}(0,0,\omega_1,\omega_2)&=&\frac{-e^3}{V}\sum_{\mathbf{k},a,b,c}\bigg[\frac{f_{ab}}{\varepsilon_{ab}-\omega_1}\langle u_b(\mathbf{k})|\partial_{\mu}\hat{v}_{\gamma}(\mathbf{k})|u_a(\mathbf{k})\rangle\langle u_a(\mathbf{k})|\hat{v}_{\nu}(\mathbf{k})|u_b(\mathbf{k})\rangle \nonumber\\&&+\frac{f_{ab}}{2(\varepsilon_{ab}-\omega)}\langle u_b(\mathbf{k})|\hat{v}_{\mu}(\mathbf{k})|u_a(\mathbf{k})\rangle\langle u_a(\mathbf{k})|\partial_{\nu}\hat{v}_{\gamma}(\mathbf{k})|u_b(\mathbf{k})\rangle\nonumber\\&&+\langle u_c(\mathbf{k})|\hat{v}_{\mu}(\mathbf{k})|u_a(\mathbf{k})\rangle\langle u_a(\mathbf{k})|\hat{v}_{\nu}(\mathbf{k})|u_b(\mathbf{k})\rangle
\langle u_b(\mathbf{k})|\hat{v}_{\gamma}(\mathbf{k})|u_c(\mathbf{k})\rangle\nonumber\\&&\times \frac{1}{\varepsilon_{ab}-\omega_1}\bigg(\frac{f_{ac}}{\varepsilon_{ac}-\omega}-\frac{f_{bc}}{\varepsilon_{bc}-\omega_2}\bigg)+(1,\nu\leftrightarrow 2,\gamma)
\bigg],
\end{eqnarray}
where $f_{ab}\equiv f(\varepsilon_a(\mathbf{k}))-f(\varepsilon_b(\mathbf{k}))$. For a two-band system, the indices $a,b,c$ can only refer to valence and conduction bands. If  these indices refers to the same band, Eq.~(\ref{46}) will be zero. Therefore,  the indices $a,b,c$ should be $a,b,a;a,b,b;a,a,b$. Taking  the derivative of Eq.~(\ref{46}) with respect to $\omega_1$ 
and exchanging the indices
$a \leftrightarrow b$ in terms containing $f(\varepsilon_b(\mathbf{k}))$, we get
\begin{eqnarray}\label{48}
\sigma^{(1,0)}_{\mu\nu\gamma}&=&\frac{-e^3}{i\omega_2V}\sum_{\mathbf{k},a\ne b}\mathrm{Im}\bigg[\frac{-2f(\varepsilon_a(\mathbf{k}))}{\varepsilon^2_{ab}}\langle u_a(\mathbf{k})|\partial_{\mu}\hat{v}_{\gamma}(\mathbf{k})|u_b(\mathbf{k})\rangle\langle u_b(\mathbf{k})|\hat{v}_{\nu}(\mathbf{k})|u_a(\mathbf{k})\rangle\nonumber\\&& -\frac{2f(\varepsilon_a(\mathbf{k}))}{\varepsilon^2_{ab}}\langle u_a(\mathbf{k})|\hat{v}_{\mu}(\mathbf{k})|u_b(\mathbf{k})\rangle\langle u_b(\mathbf{k})|\partial_{\nu}\hat{v}_{\gamma}(\mathbf{k})|u_a(\mathbf{k})\rangle\nonumber\\&&-\frac{2f(\varepsilon_a(\mathbf{k}))}{\varepsilon^3_{ab}}(v_{a\mu}-v_{b\mu})\langle u_a(\mathbf{k})|\hat{v}_{\nu}(\mathbf{k})|u_b(\mathbf{k})\rangle
\langle u_b(\mathbf{k})|\hat{v}_{\gamma}(\mathbf{k})|u_a(\mathbf{k})\rangle\nonumber\\&&
+\frac{4f(\varepsilon_a(\mathbf{k}))}{\varepsilon^3_{ab}}(v_{a\gamma}-v_{b\gamma})\langle u_a(\mathbf{k})|\hat{v}_{\mu}(\mathbf{k})|u_b(\mathbf{k})\rangle\langle u_b(\mathbf{k})|\hat{v}_{\nu}(\mathbf{k})|u_a(\mathbf{k})\rangle\nonumber\\&&
+\frac{2f(\varepsilon_a(\mathbf{k}))}{\varepsilon^3_{ab}}(v_{a\nu}-v_{b\nu})
\langle u_a(\mathbf{k})|\hat{v}_{\mu}(\mathbf{k})|u_b(\mathbf{k})\rangle\langle u_b(\mathbf{k})|\hat{v}_{\gamma}(\mathbf{k})|u_a(\mathbf{k})\rangle
\bigg]
\nonumber\\&=&\frac{ie^3}{\omega_2V}\sum_{\mathbf{k},a\ne b}f(\varepsilon_a(\mathbf{k}))\partial_{\gamma}F^{ab}_{\mu\nu}.
\end{eqnarray}
Under the interchange $1,\nu\leftrightarrow 2,\gamma$, we can also obtain  $\sigma^{(0,1)}_{\mu\nu\gamma}=(ie^3/\omega_1V)\sum_{\mathbf{k},a\ne b}f(\varepsilon_a(\mathbf{k}))\partial_{\nu}F^{ab}_{\mu\gamma}$.
In the case of SHG, i.e., $\omega_1=\omega_2=\omega_0$, Eq.~(\ref{46}) becomes
\begin{eqnarray}\label{460}
\sum^4_{i=2}\Pi^{(i)}_{\mu\nu\gamma}(0,0,\omega_0,\omega_0)&=&\frac{-e^3}{V}\sum_{\mathbf{k},a,b,c}\bigg[\frac{f_{ab}}{\varepsilon_{ab}-\omega_0}\langle u_b(\mathbf{k})|\partial_{\mu}\hat{v}_{\gamma}(\mathbf{k})|u_a(\mathbf{k})\rangle\langle u_a(\mathbf{k})|\hat{v}_{\nu}(\mathbf{k})|u_b(\mathbf{k})\rangle \nonumber\\&&+\frac{f_{ab}}{2(\varepsilon_{ab}-2\omega_0)}\langle u_b(\mathbf{k})|\hat{v}_{\mu}(\mathbf{k})|u_a(\mathbf{k})\rangle\langle u_a(\mathbf{k})|\partial_{\nu}\hat{v}_{\gamma}(\mathbf{k})|u_b(\mathbf{k})\rangle\nonumber\\&&+ \frac{1}{\varepsilon_{ab}-\omega_0}\bigg(\frac{f_{ac}}{\varepsilon_{ac}-2\omega_0}-\frac{f_{bc}}{\varepsilon_{bc}-\omega_0}\bigg)\langle u_c(\mathbf{k})|\hat{v}_{\mu}(\mathbf{k})|u_a(\mathbf{k})\rangle\langle u_a(\mathbf{k})|\hat{v}_{\nu}(\mathbf{k})|u_b(\mathbf{k})\rangle\nonumber\\&&\times
\langle u_b(\mathbf{k})|\hat{v}_{\gamma}(\mathbf{k})|u_c(\mathbf{k})\rangle+(\nu\leftrightarrow \gamma)
\bigg].
\end{eqnarray}
Taking the derivative of this equation with respect to $\omega_0$ and exchanging the indices $a\leftrightarrow b$ in terms with $f(\varepsilon_b(\mathbf{k}))$, we obtain 
\begin{eqnarray}\label{480}
\sigma^{(1)}_{\mu\nu\gamma}&=&\frac{-e^3}{i\omega_0V}\sum_{\mathbf{k},a\ne b}\mathrm{Im}\bigg[\frac{-2f(\varepsilon_a(\mathbf{k}))}{\varepsilon^2_{ab}}\langle u_a(\mathbf{k})|\partial_{\mu}\hat{v}_{\gamma}(\mathbf{k})|u_b(\mathbf{k})\rangle\langle u_b(\mathbf{k})|\hat{v}_{\nu}(\mathbf{k})|u_a(\mathbf{k})\rangle\nonumber\\&& -\frac{2f(\varepsilon_a(\mathbf{k}))}{\varepsilon^2_{ab}}\langle u_a(\mathbf{k})|\partial_{\mu}\hat{v}_{\nu}(\mathbf{k})|u_b(\mathbf{k})\rangle\langle u_b(\mathbf{k})|\hat{v}_{\gamma}(\mathbf{k})|u_a(\mathbf{k})\rangle\nonumber\\&&
-\frac{4f(\varepsilon_a(\mathbf{k}))}{\varepsilon^2_{ab}}\langle u_a(\mathbf{k})|\hat{v}_{\mu}(\mathbf{k})|u_b(\mathbf{k})\rangle\langle u_b(\mathbf{k})|\partial_{\nu}\hat{v}_{\gamma}(\mathbf{k})|u_a(\mathbf{k})\rangle\nonumber\\&&
+\frac{6f(\varepsilon_a(\mathbf{k}))}{\varepsilon^3_{ab}}(v_{a\gamma}-v_{b\gamma})\langle u_a(\mathbf{k})|\hat{v}_{\mu}(\mathbf{k})|u_b(\mathbf{k})\rangle\langle u_b(\mathbf{k})|\hat{v}_{\nu}(\mathbf{k})|u_a(\mathbf{k})\rangle\nonumber\\&&
+\frac{6f(\varepsilon_a(\mathbf{k}))}{\varepsilon^3_{ab}}(v_{a\nu}-v_{b\nu})
\langle u_a(\mathbf{k})|\hat{v}_{\mu}(\mathbf{k})|u_b(\mathbf{k})\rangle\langle u_b(\mathbf{k})|\hat{v}_{\gamma}(\mathbf{k})|u_a(\mathbf{k})\rangle
\bigg]\nonumber\\&=&
\frac{ie^3}{\omega_0V}\sum_{\mathbf{k},a\ne b}f(\varepsilon_a(\mathbf{k}))\bigg(\partial_{\gamma}F^{ab}_{\mu\nu}+\partial_{\nu}F^{ab}_{\mu\gamma}\bigg).
\end{eqnarray}
The above two types of conductivity both reproduce 
the Berry curvature dipole $\partial_{\gamma}F^{ab}_{\mu\nu}$. This quantity  has been derived from various theoretical frameworks, including the semi-classical equations of motion for an electron wave packet ~\cite{sodemann2015quantum}, the Floquet formalism ~\cite{morimoto2016semiclassical}, and the density matrix method~\cite{gao2020second}.
In the presence of time-reversal symmetry, the Berry curvature dipole is an even function of momentum. As a result, both conductivities  support time-reversal symmetry.
 Additionally, 
 the conductivity $\sigma^{(1,0)}_{\mu\nu\gamma}+\sigma^{(0,1)}_{\mu\nu\gamma}$ does not displays symmetry with respect to the indices $\nu$ and $\gamma$, whereas $\sigma^{(1)}_{\mu\nu\gamma}$ does exhibit such a symmetry.

\subsection{Nonlinear Hall effect in time-reversal breaking systems}
Consider the conductivity given by  Eq.~(\ref{45}). Taking the derivative of Eq.~(\ref{46}) with respect to $\omega_1$ and $\omega_2$  
and exchanging the indices
$a \leftrightarrow b$ in terms containing $f(\varepsilon_b(\mathbf{k}))$, we have
\begin{eqnarray}\label{50}
\sigma^{(1,1)}_{\mu\nu\gamma}&=&\frac{e^3}{V}\sum_{\mathbf{k},a\ne b}\mathrm{Re}\bigg[\frac{4f(\varepsilon_a(\mathbf{k}))}{\varepsilon^3_{ab}}\langle u_a(\mathbf{k})|\hat{v}_{\mu}(\mathbf{k})|u_b(\mathbf{k})\rangle\langle u_b(\mathbf{k})|\partial_{\nu}\hat{v}_{\gamma}(\mathbf{k})|u_a(\mathbf{k})\rangle\nonumber\\&&+\frac{2f(\varepsilon_a(\mathbf{k}))}{\varepsilon^4_{ab}}(v_{a\mu}-v_{b\mu})\langle u_a(\mathbf{k})|\hat{v}_{\nu}(\mathbf{k})|u_b(\mathbf{k})\rangle
\langle u_b(\mathbf{k})|\hat{v}_{\gamma}(\mathbf{k})|u_a(\mathbf{k})\rangle\nonumber\\&&
-\frac{6f(\varepsilon_a(\mathbf{k}))}{\varepsilon^4_{ab}}(v_{a\gamma}-v_{b\gamma})\langle u_a(\mathbf{k})|\hat{v}_{\mu}(\mathbf{k})|u_b(\mathbf{k})\rangle\langle u_b(\mathbf{k})|\hat{v}_{\nu}(\mathbf{k})|u_a(\mathbf{k})\rangle\nonumber\\&&
-\frac{6f(\varepsilon_a(\mathbf{k}))}{\varepsilon^4_{ab}}(v_{a\nu}-v_{b\nu})
\langle u_a(\mathbf{k})|\hat{v}_{\mu}(\mathbf{k})|u_b(\mathbf{k})\rangle\langle u_b(\mathbf{k})|\hat{v}_{\gamma}(\mathbf{k})|u_a(\mathbf{k})\rangle\bigg]
\nonumber\\&=&
\frac{2e^3}{V}\sum_{\mathbf{k},a\ne b}f(\varepsilon_a(\mathbf{k}))\bigg[\partial_{\nu}\bigg(\frac{g_{\mu\gamma}}{\varepsilon_{ab}}\bigg)+\partial_{\gamma}\bigg(\frac{g_{\mu\nu}}{\varepsilon_{ab}}\bigg)-\partial_{\mu}\bigg(\frac{g_{\gamma\nu}}{\varepsilon_{ab}}\bigg)\bigg],
\end{eqnarray}
where $g_{\mu\gamma}=\mathrm{Re}\langle u_a(\mathbf{k})|\hat{v}_{\mu}(\mathbf{k})|u_b(\mathbf{k})\rangle\langle u_b(\mathbf{k})|\hat{v}_{\gamma}(\mathbf{k})|u_a(\mathbf{k})\rangle/\varepsilon^2_{ab}$ is the quantum metric ~\cite{provost1980riemannian,zhang2022revealing}.
In the SHG case, we take the second order derivative of Eq.~(\ref{460})
with respect to $\omega_0$ and get the conductivity in Eq.~(\ref{45a})
\begin{eqnarray}\label{51}
\sigma^{(2)}_{\mu\nu\gamma}&=&\frac{e^3}{V}\sum_{\mathbf{k},a\ne b}\mathrm{Re}\bigg[
\frac{2f(\varepsilon_a(\mathbf{k}))}{\varepsilon^3_{ab}}\langle u_a(\mathbf{k})|\hat{v}_{\nu}(\mathbf{k})|u_b(\mathbf{k})\rangle\langle u_b(\mathbf{k})|\partial_{\mu}\hat{v}_{\gamma}(\mathbf{k})|u_a(\mathbf{k})\rangle
\nonumber\\&&+\frac{2f(\varepsilon_a(\mathbf{k}))}{\varepsilon^3_{ab}}\langle u_a(\mathbf{k})|\hat{v}_{\gamma}(\mathbf{k})|u_b(\mathbf{k})\rangle\langle u_b(\mathbf{k})|\partial_{\mu}\hat{v}_{\nu}(\mathbf{k})|u_a(\mathbf{k})\rangle\nonumber\\&&+
\frac{8f(\varepsilon_a(\mathbf{k}))}{\varepsilon^3_{ab}}\langle u_a(\mathbf{k})|\hat{v}_{\mu}(\mathbf{k})|u_b(\mathbf{k})\rangle\langle u_b(\mathbf{k})|\partial_{\nu}\hat{v}_{\gamma}(\mathbf{k})|u_a(\mathbf{k})\rangle
\nonumber\\&&-\frac{2f(\varepsilon_a(\mathbf{k}))}{\varepsilon^4_{ab}}(v_{a\mu}-v_{b\mu})\langle u_a(\mathbf{k})|\hat{v}_{\nu}(\mathbf{k})|u_b(\mathbf{k})\rangle
\langle u_b(\mathbf{k})|\hat{v}_{\gamma}(\mathbf{k})|u_a(\mathbf{k})\rangle\nonumber\\&&
-\frac{14f(\varepsilon_a(\mathbf{k}))}{\varepsilon^4_{ab}}(v_{a\gamma}-v_{b\gamma})\langle u_a(\mathbf{k})|\hat{v}_{\mu}(\mathbf{k})|u_b(\mathbf{k})\rangle\langle u_b(\mathbf{k})|\hat{v}_{\nu}(\mathbf{k})|u_a(\mathbf{k})\rangle
\nonumber\\&&-\frac{14f(\varepsilon_a(\mathbf{k}))}{\varepsilon^4_{ab}}(v_{a\nu}-v_{b\nu})
\langle u_a(\mathbf{k})|\hat{v}_{\mu}(\mathbf{k})|u_b(\mathbf{k})\rangle\langle u_b(\mathbf{k})|\hat{v}_{\gamma}(\mathbf{k})|u_a(\mathbf{k})\rangle\bigg]\nonumber\\&=&
\frac{2e^3}{V}\sum_{\mathbf{k},a\ne b}f(\varepsilon_a(\mathbf{k}))\bigg[2\partial_{\nu}\bigg(\frac{g_{\mu\gamma}}{\varepsilon_{ab}}\bigg)+2\partial_{\gamma}\bigg(\frac{g_{\mu\nu}}{\varepsilon_{ab}}\bigg)-\partial_{\mu}\bigg(\frac{g_{\gamma\nu}}{\varepsilon_{ab}}\bigg)\bigg].
\end{eqnarray}
These two types of conductivity mentioned above produce different values, but they are both closely related to 
the band-energy normalized quantum metric  $2g_{\mu\nu}/\varepsilon_{ab}$    and the normalized quantum
metric dipole $\partial_{\gamma}(2g_{\mu\nu}/\varepsilon_{ab})$ ~\cite{wang2023quantum}. Our results demonstrate how the nonlinear Hall effects manifest as the frequencies of the fields approach zero along different pathways.
In the presence of time-reversal symmetry, both the energy dispersion and the quantum metric are even functions of momentum. As a result,  Eq.~(\ref{50}) and Eq.~(\ref{51}) vanish. Therefore, these two  conductivities  require the breaking of time-reversal symmetry.
Besides, these two nonlinear responses are intrinsic and symmetric under the interchange  $\nu\leftrightarrow\gamma$.

\section{Magneto-nonlinear Hall effects}
Next, we explore the magneto-nonlinear Hall effects in both time-reversal symmetric and time-
reversal breaking systems,
with a particular focus on its manifestation in  SHG case.

From  Eqs.~(\ref{25}) to (\ref{32}), it can be seen that the current in SHG case can be formulated as
\begin{eqnarray}\label{36}
\langle \hat{j}_{\mu}(2\mathbf{q}_0,2i\omega_0) \rangle=\sum^4_{i=1}\Pi^{(i)}_{\mu\nu\gamma}(\mathbf{q}_0,\mathbf{q}_0,\omega_0,\omega_0)A_{\nu}(\mathbf{q}_0,\omega_0)A_{\gamma}(\mathbf{q}_0,\omega_0).
\end{eqnarray}
To derive the response to the electric and magnetic fields $E_{\nu}(\mathbf{q}_0,\omega_0)B_{\rho}(\mathbf{q}_0,\omega_0)$,
we need to expand the 
 current to  first order in $\mathbf{q}_0$. Note that the  coefficient $\Pi^{(1)}_{\mu\nu\gamma}(\mathbf{q}_0,\mathbf{q}_0,0,0)$ in Eq.~(\ref{3001z}) is an even function of $\mathbf{q}_0$, this term does not contribute to such a response. Consequently, we have
\begin{eqnarray}\label{36g}
\langle \hat{j}_{\mu}(2\mathbf{q}_0,2i\omega_0) \rangle&=&\sum^4_{i=2}\partial_{{q_{0}}_\eta}\Pi^{(i)}_{\mu\nu\gamma}(\mathbf{q}_0,\mathbf{q}_0,\omega_0,\omega_0)|_{{{q_{0}}_\eta}=0}{q_{0}}_\eta A_{\nu}(\mathbf{q}_0,\omega_0)A_{\gamma}(\mathbf{q}_0,\omega_0)\nonumber\\&\equiv &\alpha_{\mu\gamma\rho}(\omega_0) E_{\gamma}(\mathbf{q}_0,\omega_0)B_{\rho}(\mathbf{q}_0,\omega_0),
\end{eqnarray}
 where
 \begin{eqnarray}\label{36a}
 \alpha_{\mu\gamma\rho}(\omega_0)=-\frac{\epsilon_{\rho\eta\nu}}{\omega_0}\sum^4_{i=2}\partial_{{q_{0}}_\eta}\Pi^{(i)}_{\mu\nu\gamma}(\mathbf{q}_0,\mathbf{q}_0,\omega_0,\omega_0)|_{{{q_{0}}_\eta}=0}
 \end{eqnarray}
 is the response coefficient, $B_{\rho}(\mathbf{q}_0,\omega_0)=i\epsilon_{\rho\eta\nu}{q_{0}}_\eta A_{\nu}(\mathbf{q}_0,\omega_0)$, and $\eta$-axis is the direction of the the wave vector $\mathbf{q}_0$.
By differentiating  Eq.~(\ref{34})  in the SHG case with respect to  $\mathbf{q}_0$, we can obtain the relation
\begin{eqnarray}\label{36}
\partial_{{q_{0}}_\eta}\Pi^{(i)}_{\mu\nu\gamma}(\mathbf{q}_0,\mathbf{q}_0,\omega_0,\omega_0)|_{{{q_{0}}_\eta}=0}=-\partial_{{q_{0}}_\eta}\Pi^{(i)*}_{\mu\nu\gamma}(\mathbf{q}_0,\mathbf{q}_0,-\omega_0,-\omega_0)|_{{{q_{0}}_\eta}=0}.
\end{eqnarray}
We perform a Taylor expansion of   the above relation to first order in $\omega_0$, compare the coefficients at the zeroth and first order, and have
\begin{eqnarray}\label{37t}
\partial_{{q_{0}}_\eta}\Pi^{(i)}_{\mu\nu\gamma}(\mathbf{q}_0,\mathbf{q}_0,\omega_0,\omega_0)|_{{{q_{0}}_\eta}=0}&=&\partial_{{q_{0}}_\eta}\Pi^{(i)}_{\mu\nu\gamma}(\mathbf{q}_0,\mathbf{q}_0,0,0)|_{{{q_{0}}_\eta}=0}\nonumber\\&&+\partial_{\omega_0}\partial_{{q_{0}}_\eta}\Pi^{(i)}_{\mu\nu\gamma}(\mathbf{q}_0,\mathbf{q}_0,\omega_0,\omega_0)|_{{{q_{0}}_\eta},\omega_0=0}\omega_0,
\end{eqnarray}
\begin{eqnarray}\label{37}
\partial_{{q_{0}}_\eta}\Pi^{(i)}_{\mu\nu\gamma}(\mathbf{q}_0,\mathbf{q}_0,0,0)|_{{{q_{0}}_\eta}=0}&=&-\partial_{{q_{0}}_\eta}\Pi^{(i)*}_{\mu\nu\gamma}(\mathbf{q}_0,\mathbf{q}_0,0,0)|_{{{q_{0}}_\eta}=0}\nonumber\\&=&i\mathrm{Im}\partial_{{q_{0}}_\eta}\Pi^{(i)}_{\mu\nu\gamma}(\mathbf{q}_0,\mathbf{q}_0,0,0)|_{{{q_{0}}_\eta}=0},
\end{eqnarray}
and
\begin{eqnarray}\label{37a}
\partial_{\omega_0}\partial_{{q_{0}}_\eta}\Pi^{(i)}_{\mu\nu\gamma}(\mathbf{q}_0,\mathbf{q}_0,\omega_0,\omega_0)|_{{{q_{0}}_\eta},\omega_0=0}&=&\partial_{\omega_0}\partial_{{q_{0}}_\eta}\Pi^{(i)*}_{\mu\nu\gamma}(\mathbf{q}_0,\mathbf{q}_0,\omega_0,\omega_0)|_{{{q_{0}}_\eta},\omega_0=0}\nonumber\\&=&\mathrm{Re}\partial_{\omega_0}\partial_{{q_{0}}_\eta}\Pi^{(i)}_{\mu\nu\gamma}(\mathbf{q}_0,\mathbf{q}_0,\omega_0,\omega_0)|_{{{q_{0}}_\eta},\omega_0=0}.
\end{eqnarray}
Substituting Eq.~(\ref{37t}) into the expression for  the response coefficient, namely  Eq.~(\ref{36a}), and utilizing  the relations in Eqs.~(\ref{37}) and ~(\ref{37a}), we have two types of response coefficient. The one generated by the zeroth-order term of  Eq.~(\ref{37t}) is  given by
\begin{eqnarray}\label{38a}
\alpha^{(0)}_{\mu\gamma\rho}=\frac{\epsilon_{\rho\eta\nu}}{i\omega_0}\sum^4_{i=2}\mathrm{Im}\partial_{{q_{0}}_\eta}\Pi^{(i)}_{\mu\nu\gamma}(\mathbf{q}_0,\mathbf{q}_0,0,0)|_{{{q_{0}}_\eta}=0},
\end{eqnarray}
which exhibits a dependence on the frequency of the external fields.
Meanwhile, the one derived from the first-order term of  Eq.~(\ref{37t}) is  expressed as
\begin{eqnarray}\label{38b}
\alpha^{(1)}_{\mu\gamma\rho}=-\epsilon_{\rho\eta\nu}\sum^4_{i=2}\mathrm{Re}\partial_{\omega_0}\partial_{{q_{0}}_\eta}\Pi^{(i)}_{\mu\nu\gamma}(\mathbf{q}_0,\mathbf{q}_0,\omega_0,\omega_0)|_{{{q_{0}}_\eta},\omega_0=0}.
\end{eqnarray}
This response is independent of the frequency  as well as the relaxation time, thus  it represents an intrinsic response. 

\subsection{Magneto-nonlinear Hall effect in time-reversal breaking systems}
Consider the response given by in Eq.~(\ref{38a}). For simplicity, 
we consider a  linear continuum Hamiltonian targeting various Dirac and Weyl systems. This means that the Hamiltonian of the sample satisfies the restriction:  $\partial_{\mu} \partial_{\nu}H(\mathbf{k})=\partial_{\mu} \hat{v}_{\nu}(\mathbf{k})=0$. Under such condition,
we have
\begin{eqnarray}\label{52}
\sum^4_{i=2}\Pi^{(i)}_{\mu\nu\gamma}(\mathbf{q}_0,\mathbf{q}_0,\omega_0,\omega_0)&=&\frac{-e^3}{V}\sum_{\mathbf{k},a, b,c}\bigg\{\frac{1}{\varepsilon_a(\mathbf{k})-\varepsilon_b(\mathbf{k}-\mathbf{q}_0)-\omega_0}\bigg[\frac{f(\varepsilon_a(\mathbf{k}))-f(\varepsilon_c(\mathbf{k}-2\mathbf{q}_0))}{\varepsilon_a(\mathbf{k})-\varepsilon_c(\mathbf{k}-2\mathbf{q}_0)-2\omega_0}\nonumber\\&&-\frac{f(\varepsilon_b(\mathbf{k}-\mathbf{q}_0))-f(\varepsilon_c(\mathbf{k}-2\mathbf{q}_0))}{\varepsilon_b(\mathbf{k}-\mathbf{q}_0)-\varepsilon_c(\mathbf{k}-2\mathbf{q}_0)-\omega_0}\bigg]\langle u_c(\mathbf{k}-2\mathbf{q}_0)|\hat{v}_{\mu}(\mathbf{k})|u_a(\mathbf{k})\rangle\nonumber\\&&\times\langle u_a(\mathbf{k})|\hat{v}_{\nu}(\mathbf{k})|u_b(\mathbf{k}-\mathbf{q}_0)\rangle
\langle u_b(\mathbf{k}-\mathbf{q}_0)|\hat{v}_{\gamma}(\mathbf{k})|u_c(\mathbf{k}-2\mathbf{q}_0)\rangle+(\nu\leftrightarrow\gamma)\bigg\}.\nonumber\\
\end{eqnarray}
For a two-band system, when the indices $a, b, c$ refer to the same band, $\partial_{{q_{0}}_\eta}\mathrm{Im}\Pi^{(i)}_{\mu\nu\gamma}(\mathbf{q}_0,\mathbf{q}_0,\omega_0,\omega_0)|_{{{q_{0}}_\eta}=0}$ 
will be zero. Therefore, the indices should also be $a,b,a;a,b,b;a,a,b$. 
Eq.~(\ref{52}) can be decomposed into intraband and interband parts as $\sum^4_{i=2}\Pi^{(i)}_{\mu\nu\gamma}(\mathbf{q}_0,\mathbf{q}_0,\omega_0,\omega_0)=\Pi^{intra}_{\mu\nu\gamma}(\mathbf{q}_0,\omega_0)+\Pi^{inter}_{\mu\nu\gamma}(\mathbf{q}_0,\omega_0)$. The intraband part, which corresponds to transitions within the same band but at different points in momentum space, takes the form 
\begin{eqnarray}\label{53}
\Pi^{intra}_{\mu\nu\gamma}(\mathbf{q}_0,\omega_0)&=&\frac{-e^3}{V}\sum_{\mathbf{k},a\ne b}\bigg[\frac{1}{\varepsilon_{a}(\mathbf{k})-\varepsilon_{b}(\mathbf{k}-\mathbf{q}_0)-\omega_0}\frac{f(\varepsilon_a(\mathbf{k}))-f(\varepsilon_a(\mathbf{k}-2\mathbf{q}_0))}{\varepsilon_{a}(\mathbf{k})-\varepsilon_{a}(\mathbf{k}-2\mathbf{q}_0)-2\omega_0}\nonumber\\&&\times\langle u_a(\mathbf{k}-2\mathbf{q}_0)|\hat{v}_{\mu}(\mathbf{k})|u_a(\mathbf{k})\rangle\langle u_a(\mathbf{k})|\hat{v}_{\nu}(\mathbf{k})|u_b(\mathbf{k}-\mathbf{q}_0)\rangle
\langle u_b(\mathbf{k}-\mathbf{q}_0)|\hat{v}_{\gamma}(\mathbf{k})|u_a(\mathbf{k}-2\mathbf{q}_0)\rangle\nonumber\\&&+\frac{-1}{\varepsilon_{a}(\mathbf{k})-\varepsilon_{b}(\mathbf{k}-\mathbf{q}_0)-\omega_0}\frac{f(\varepsilon_b(\mathbf{k}-\mathbf{q}_0))-f(\varepsilon_b(\mathbf{k}-2\mathbf{q}_0))}{\varepsilon_b(\mathbf{k}-\mathbf{q}_0)-\varepsilon_b(\mathbf{k}-2\mathbf{q}_0)-\omega_0}\nonumber\\&&\times\langle u_b(\mathbf{k}-2\mathbf{q}_0)|\hat{v}_{\mu}(\mathbf{k})|u_a(\mathbf{k})\rangle\langle u_a(\mathbf{k})|\hat{v}_{\nu}(\mathbf{k})|u_b(\mathbf{k}-\mathbf{q}_0)\rangle
\langle u_b(\mathbf{k}-\mathbf{q}_0)|\hat{v}_{\gamma}(\mathbf{k})|u_b(\mathbf{k}-2\mathbf{q}_0)\rangle\nonumber\\&&+\frac{1}{\varepsilon_{a}(\mathbf{k})-\varepsilon_{b}(\mathbf{k}-2\mathbf{q}_0)-2\omega_0}\frac{f(\varepsilon_a(\mathbf{k}))-f(\varepsilon_a(\mathbf{k}-\mathbf{q}_0))}{\varepsilon_{a}(\mathbf{k})-\varepsilon_{a}(\mathbf{k}-\mathbf{q}_0)-\omega_0}\nonumber\\&&\times\langle u_b(\mathbf{k}-2\mathbf{q}_0)|\hat{v}_{\mu}(\mathbf{k})|u_a(\mathbf{k})\rangle\langle u_a(\mathbf{k})|\hat{v}_{\nu}(\mathbf{k})|u_a(\mathbf{k}-\mathbf{q}_0)\rangle
\langle u_a(\mathbf{k}-\mathbf{q}_0)|\hat{v}_{\gamma}(\mathbf{k})|u_b(\mathbf{k}-2\mathbf{q}_0)\rangle\nonumber\\&&+(\nu\leftrightarrow\gamma)\bigg].
\end{eqnarray}
Upon differentiating   Eq.~(\ref{53}) with respect to ${q_{0}}_\eta$ and   subsequently swapping the indices $a, b$ in the terms involving $f(\varepsilon_b(\mathbf{k}))$, one can find that under the interchange $\nu\leftrightarrow\gamma$, the first term in Eq.~(\ref{53})  yields  a
 real number which does not contribute to the response. Therefore one gets
\begin{eqnarray}\label{54}
\partial_{{q_{0}}_\eta}\Pi^{intra}_{\mu\nu\gamma}(\mathbf{q}_0,\omega_0)|_{{{q_{0}}_\eta}=0}&=&\frac{-e^3}{V}\sum_{\mathbf{k},a\ne b}\bigg[\frac{-f^{'}(\varepsilon_a(\mathbf{k}))}{(\varepsilon_{ab}+\omega_0)\omega_0}v_{a\gamma}\langle u_a(\mathbf{k})|\hat{v}_{\mu}(\mathbf{k})|u_b(\mathbf{k})\rangle\langle u_b(\mathbf{k})|\hat{v}_{\nu}(\mathbf{k})|u_a(\mathbf{k})\rangle
 \nonumber\\&&+\frac{-f^{'}(\varepsilon_a(\mathbf{k}))}{(\varepsilon_{ab}-2\omega_0)\omega_0}v_{a\nu}\langle u_b(\mathbf{k})|\hat{v}_{\mu}(\mathbf{k})|u_a(\mathbf{k})\rangle\rangle\langle u_a(\mathbf{k})|\hat{v}_{\gamma}(\mathbf{k})|u_b(\mathbf{k})\rangle+(\nu\leftrightarrow\gamma)\bigg],\nonumber\\
\end{eqnarray}
where $f^{'}(\varepsilon_a(\mathbf{k}))$ refers to $\partial_{k_{\eta}}f(\varepsilon_a(\mathbf{k}))$.
In the limit $\omega_0\rightarrow0$, we have
\begin{eqnarray}\label{540}
\mathrm{Im}\partial_{{q_{0}}_\eta}\Pi^{intra}_{\mu\nu\gamma}(\mathbf{q}_0,0)|_{{{q_{0}}_\eta}=0}&=&\frac{-e^3}{V}\sum_{\mathbf{k},a\ne b}\bigg[\frac{3f^{'}(\varepsilon_a(\mathbf{k}))}{\varepsilon^2_{ab}}\mathrm{Im}\bigg(v_{a\gamma}\langle u_a(\mathbf{k})|\hat{v}_{\mu}(\mathbf{k})|u_b(\mathbf{k})\rangle\langle u_b(\mathbf{k})|\hat{v}_{\nu}(\mathbf{k})|u_a(\mathbf{k})\rangle 
\nonumber\\&&+v_{a\nu}\langle u_a(\mathbf{k})|\hat{v}_{\mu}(\mathbf{k})|u_b(\mathbf{k})\rangle\langle u_b(\mathbf{k})|\hat{v}_{\gamma}(\mathbf{k})|u_a(\mathbf{k})\rangle \bigg)\bigg].
\end{eqnarray}

 The interband part  corresponds to transitions between different bands and  is given by
\begin{eqnarray}\label{55}
\Pi^{inter}_{\mu\nu\gamma}(\mathbf{q}_0,0)&=&\frac{-e^3}{V}\sum_{\mathbf{k},a\ne b}\bigg[\frac{-1}{\varepsilon_{a}(\mathbf{k})-\varepsilon_{b}(\mathbf{k}-\mathbf{q}_0)-\omega_0}\frac{f(\varepsilon_a(\mathbf{k}-2\mathbf{q}_0))-f(\varepsilon_b(\mathbf{k}-\mathbf{q}_0))}{\varepsilon_{a}(\mathbf{k}-2\mathbf{q}_0)-\varepsilon_{b}(\mathbf{k}-\mathbf{q}_0)+\omega_0}\nonumber\\&&\times\langle u_a(\mathbf{k}-2\mathbf{q}_0)|\hat{v}_{\mu}(\mathbf{k})|u_a(\mathbf{k})\rangle\langle u_a(\mathbf{k})|\hat{v}_{\nu}(\mathbf{k})|u_b(\mathbf{k}-\mathbf{q}_0)\rangle
\langle u_b(\mathbf{k}-\mathbf{q}_0)|\hat{v}_{\gamma}(\mathbf{k})|u_a(\mathbf{k}-2\mathbf{q}_0)\rangle\nonumber\\&&+\frac{1}{\varepsilon_{a}(\mathbf{k})-\varepsilon_{b}(\mathbf{k}-\mathbf{q}_0)-\omega_0}\frac{f(\varepsilon_a(\mathbf{k}))-f(\varepsilon_b(\mathbf{k}-2\mathbf{q}_0))}{\varepsilon_a(\mathbf{k})-\varepsilon_b(\mathbf{k}-2\mathbf{q}_0)-2\omega_0}\nonumber\\&&\times\langle u_b(\mathbf{k}-2\mathbf{q}_0)|\hat{v}_{\mu}(\mathbf{k})|u_a(\mathbf{k})\rangle\langle u_a(\mathbf{k})|\hat{v}_{\nu}(\mathbf{k})|u_b(\mathbf{k}-\mathbf{q}_0)\rangle
\langle u_b(\mathbf{k}-\mathbf{q}_0)|\hat{v}_{\gamma}(\mathbf{k})|u_b(\mathbf{k}-2\mathbf{q}_0)\rangle\nonumber\\&&+\frac{-1}{\varepsilon_{a}(\mathbf{k})-\varepsilon_{b}(\mathbf{k}-2\mathbf{q}_0)-2\omega_0}\frac{f(\varepsilon_a(\mathbf{k}-\mathbf{q}_0))-f(\varepsilon_b(\mathbf{k}-2\mathbf{q}_0))}{\varepsilon_{a}(\mathbf{k}-\mathbf{q}_0)-\varepsilon_{b}(\mathbf{k}-2\mathbf{q}_0)-\omega_0}\nonumber\\&&\times\langle u_b(\mathbf{k}-2\mathbf{q}_0)|\hat{v}_{\mu}(\mathbf{k})|u_a(\mathbf{k})\rangle\langle u_a(\mathbf{k})|\hat{v}_{\nu}(\mathbf{k})|u_a(\mathbf{k}-\mathbf{q}_0)\rangle
\langle u_a(\mathbf{k}-\mathbf{q}_0)|\hat{v}_{\gamma}(\mathbf{k})|u_b(\mathbf{k}-2\mathbf{q}_0)\rangle\nonumber\\&&+(\nu\leftrightarrow\gamma)
\bigg].
\end{eqnarray}
Due to the fact that the interband part is independent  of the order of the limit $\mathbf{q}_0\rightarrow0$ and $\omega_0\rightarrow0$, we have taken $\omega_0\rightarrow0$ before $\mathbf{q}_0\rightarrow0$. Taking the derivative of Eq.~(\ref{55})  with respect to ${q_{0}}_\eta$ and exchanging the indices $a \leftrightarrow b$ in terms containing $f(\varepsilon_b(\mathbf{k}))$, one obtains
\begin{eqnarray}\label{56}
\mathrm{Im}\partial_{{q_{0}}_\eta}\Pi^{inter}_{\mu\nu\gamma}(\mathbf{q}_0,0)|_{{{q_{0}}_\eta}=0}&=&\frac{-e^3}{V}\sum_{\mathbf{k},a\ne b}\mathrm{Im}\bigg\{\frac{f^{'}(\varepsilon_a(\mathbf{k}))}{\varepsilon^2_{ab}} \langle u_a(\mathbf{k})|\hat{v}_{\mu}(\mathbf{k})|u_b(\mathbf{k})\rangle\langle u_b(\mathbf{k})|\hat{v}_{\gamma}(\mathbf{k})|u_a(\mathbf{k})\rangle(v_{a\nu}-2v_{b\nu})\nonumber\\&&+\frac{2f(\varepsilon_a(\mathbf{k}))}{\varepsilon^3_{ab}}\bigg[\langle u_a(\mathbf{k})|\hat{v}_{\mu}(\mathbf{k})|u_b(\mathbf{k})\rangle\langle u_b(\mathbf{k})|\hat{v}_{\gamma}(\mathbf{k})|u_a(\mathbf{k})\rangle\nonumber\\&&\times(v_{b\eta}v_{b\nu}-2v_{b\eta}v_{a\nu}-v_{a\eta}v_{a\nu}+2v_{a\eta}v_{b\nu})
\nonumber\\&&+
\langle u_a(\mathbf{k})|\hat{v}_{\gamma}(\mathbf{k})|u_b(\mathbf{k})\rangle\langle u_b(\mathbf{k})|\hat{v}_{\eta}(\mathbf{k})|u_a(\mathbf{k})\rangle(v_{b\mu}v_{b\nu}-2v_{b\mu}v_{a\nu}-v_{a\mu}v_{a\nu}+2v_{a\mu}v_{b\nu})
\nonumber\\&&+\langle u_a(\mathbf{k})|\hat{v}_{\mu}(\mathbf{k})|u_b(\mathbf{k})\rangle\langle u_b(\mathbf{k})|\hat{v}_{\eta}(\mathbf{k})|u_a(\mathbf{k})\rangle(v_{a\nu}v_{a\gamma}-v_{b\nu}v_{b\gamma})\bigg]+(\nu\leftrightarrow\gamma)\bigg\}.
\end{eqnarray}
Combining the contribution of the intraband and interband terms, the response coefficient becomes 
\begin{eqnarray}\label{56}
\alpha^{(0)}_{\mu\gamma\rho}&=&\frac{e^3\epsilon_{\rho\eta\nu}}{i\omega_0 V}\sum_{\mathbf{k},a\ne b}\bigg\{f^{'}(\varepsilon_a(\mathbf{k}))F^{ab}_{\mu\gamma}(2v_{a\nu}-v_{b\nu})\nonumber\\&&+\frac{f(\varepsilon_a(\mathbf{k}))}{\varepsilon_{ab}}\bigg[F^{ab}_{\mu\gamma}(v_{b\eta}v_{b\nu}-2v_{b\eta}v_{a\nu}-v_{a\eta}v_{a\nu}+2v_{a\eta}v_{b\nu})
-(\eta\leftrightarrow\mu)-(\eta\leftrightarrow\gamma)
\bigg]+(\nu\leftrightarrow\gamma)\bigg\}\nonumber\\&=&\frac{ie^3\epsilon_{\rho\eta\nu}}{\omega_0 V}\sum_{\mathbf{k},a\ne b}f(\varepsilon_a(\mathbf{k}))\bigg\{\partial_{\eta}\big[F^{ab}_{\mu\gamma}(2v_{a\nu}-v_{b\nu})\big]+\frac{\epsilon_{\eta\gamma\mu}}{\varepsilon_{ab}}F^{ab}_{\lambda}\big[v_{b\lambda}(v_{b\nu}-2v_{a\nu})-v_{a\lambda}(v_{a\nu}-2v_{b\nu})\big]
\nonumber\\&&+(\nu\leftrightarrow\gamma)\bigg\}.
\end{eqnarray}
Under time-reversal symmetry, both the Berry curvature and the  group velocity are odd functions of momentum, leading to a vanishing response. Consequently, such a response  requires the breaking of time-reversal symmetry.
The quantity $\partial_{\eta}\big(F^{ab}_{\mu\gamma}v_{a\nu}\big)$ in the first term can be regarded as the Berry curvature-velocity  dipole.  In a two-dimensional system, the  fact $F^{ab}_{\lambda}v_{i\lambda}=0$  ensures that the second term  in Eq.~(\ref{56}) vanishes. 
When the current flows in the x-direction, the electric field is polarized in the $y$-direction and the magnetic field in the $z$-direction, both propagating along the $x$-direction. Specifically, we have   $\mu=x,\nu=y,\gamma=y,\eta=x$, and $\rho=z$. Under these conditions, the above equation  reduces to the result in  Ref.~\cite{zhang2023nonlinear}.

\subsection{Magneto-nonlinear Hall effect in time-reversal symmetric systems}
Now consider the response in  Eq.~(\ref{38b}). Taking the derivative of Eq.~(\ref{52})  with respect to ${q_{0}}_\eta$ and $\omega_0$, the intraband part gives
\begin{eqnarray}\label{57}
\mathrm{Re}\partial_{\omega_0}\partial_{{q_{0}}_\eta}\Pi^{intra}_{\mu\nu\gamma}(\mathbf{q}_0,\omega_0)|_{{{q_{0}}_\eta},\omega_0=0}&=&\frac{-e^3}{V}\sum_{\mathbf{k},a\ne b}\frac{f^{'}(\varepsilon_a(\mathbf{k}))}{\varepsilon_{ab}}(-2g_{\nu\gamma}v_{a\mu}-5g_{\mu\gamma}v_{a\nu}-5g_{\mu\nu}v_{a\gamma}). 
\end{eqnarray}
And the interband term yields
\begin{eqnarray}\label{58}
\mathrm{Re}\partial_{\omega_0}\partial_{{q_{0}}_\eta}\Pi^{inter}_{\mu\nu\gamma}(\mathbf{q}_0,\omega_0)|_{{{q_{0}}_\eta},\omega_0=0}&=&\frac{-e^3}{V}\sum_{\mathbf{k},a\ne b}\bigg\{\frac{3f^{'}(\varepsilon_a(\mathbf{k}))}{\varepsilon_{ab}}\big[g_{\mu\gamma}(2v_{b\nu}-v_{a\nu})+g_{\mu\nu}(2v_{b\gamma}-v_{a\gamma})\big]\nonumber\\&&+2f(\varepsilon_a(\mathbf{k}))(v_{a\mu}+v_{b\mu})\bigg[\partial_{\nu}\bigg(\frac{g_{\eta\gamma}}{\varepsilon_{ab}}\bigg)+\partial_{\gamma}\bigg(\frac{g_{\eta\nu}}{\varepsilon_{ab}}\bigg)-\partial_{\eta}\bigg(\frac{g_{\gamma\nu}}{\varepsilon_{ab}}\bigg)\bigg]\nonumber\\&&-2f(\varepsilon_a(\mathbf{k}))(v_{a\eta}+v_{b\eta})\bigg[2\partial_{\nu}\bigg(\frac{g_{\mu\gamma}}{\varepsilon_{ab}}\bigg)+2\partial_{\gamma}\bigg(\frac{g_{\mu\nu}}{\varepsilon_{ab}}\bigg)-\partial_{\mu}\bigg(\frac{g_{\gamma\nu}}{\varepsilon_{ab}}\bigg)\bigg]\nonumber\\&&+f(\varepsilon_a(\mathbf{k}))(v_{a\nu}+v_{b\nu})\bigg[2\partial_{\gamma}\bigg(\frac{g_{\mu\eta}}{\varepsilon_{ab}}\bigg)+\partial_{\eta}\bigg(\frac{g_{\mu\gamma}}{\varepsilon_{ab}}\bigg)-\partial_{\mu}\bigg(\frac{g_{\gamma\eta}}{\varepsilon_{ab}}\bigg)\bigg]\nonumber\\&&+f(\varepsilon_a(\mathbf{k}))(v_{a\gamma}+v_{b\gamma})\bigg[2\partial_{\nu}\bigg(\frac{g_{\mu\eta}}{\varepsilon_{ab}}\bigg)+\partial_{\eta}\bigg(\frac{g_{\mu\nu}}{\varepsilon_{ab}}\bigg)-\partial_{\mu}\bigg(\frac{g_{\nu\eta}}{\varepsilon_{ab}}\bigg)\bigg]\bigg\}. \nonumber\\
\end{eqnarray}
Here,  Eq.~(\ref{50}) and  Eq.~(\ref{51}) are used. Combining the contributions of the intraband and interband terms, the response coefficient  becomes 
\begin{eqnarray}\label{59}
\alpha^{(1)}_{\mu\gamma\rho}&=&\frac{-e^3\epsilon_{\rho\eta\nu}}{V}\sum_{\mathbf{k},a\ne b}f(\varepsilon_a(\mathbf{k}))\bigg\{2\partial_{\nu}\bigg[\frac{g_{\eta\gamma}}{\varepsilon_{ab}}(v_{a\mu}+v_{b\mu})+\frac{g_{\mu\eta}}{\varepsilon_{ab}}(v_{a\gamma}+v_{b\gamma})-2\frac{g_{\mu\gamma}}{\varepsilon_{ab}}(v_{a\eta}+v_{b\eta})\bigg]\nonumber\\&&+\partial_{\mu}\bigg[\frac{g_{\nu\gamma}}{\varepsilon_{ab}}(v_{a\eta}+v_{b\eta})-\frac{g_{\eta\gamma}}{\varepsilon_{ab}}(v_{a\nu}+v_{b\nu})\bigg]+\partial_{\eta}\bigg[\frac{g_{\mu\gamma}}{\varepsilon_{ab}}(9v_{a\nu}-5v_{b\nu})-\frac{g_{\nu\gamma}}{\varepsilon_{ab}}v_{b\mu}\bigg]
+(\nu\leftrightarrow\gamma)\bigg\}. \nonumber\\
\end{eqnarray}
In the presence of time-reversal symmetry, the quantum metric and energy dispersion are even functions of momentum, whereas the group velocity is an odd function of momentum. These   properties ensure that this magneto-nonlinear Hall effect  can manifest in systems with time-reversal symmetry. However, it is worth noting that this does not mean that this response vanishes in time-reversal breaking systems.  In fact, it can also exist in systems that break time-reversal symmetry, such as  massive Dirac model.
For a two-band  model without an overall energy shift, the quantity $v_{a\alpha}+v_{b\alpha}=\partial_{\alpha}(\varepsilon_a+\varepsilon_b)$ vanishes. As a result, this transport coefficient is significantly simplified to
\begin{eqnarray}\label{60}
\alpha^{(1)}_{\mu\gamma\rho}=\frac{-e^3\epsilon_{\rho\eta\nu}}{V}\sum_{\mathbf{k},a\ne b}f(\varepsilon_a(\mathbf{k}))\partial_{\eta}\bigg[\frac{7g_{\mu\gamma}}{\varepsilon_{ab}}(v_{a\nu}-v_{b\nu})-\frac{g_{\nu\gamma}}{\varepsilon_{ab}}v_{b\mu}\bigg]
+(\nu\leftrightarrow\gamma).
\end{eqnarray}
The quantity $\partial_{\eta}\big(\frac{2g_{\mu\gamma}}{\varepsilon_{ab}}v_{a\nu}\big)$ in the above equations can be deemed as the normalized quantum metric-velocity dipole.
\section{The impact of Zeeman interaction}
In the previous sections, the spin of the electron was not taken into account. If the spin is considered, the coupling between the magnetic and the spin will gives rise to a Zeeman interaction term  in the interaction Hamiltonian
\begin{eqnarray}\label{57z}
H^{'}_{ze}=g\mu_B \mathbf{B}\cdot\mathbf{S},
\end{eqnarray}
where $g$ is the electron $g$ factor, $\mu_B$ is the Bohr magneton, and $\mathbf{S}=\mathbf{\sigma}/2$ is the spin operator. 
The Zeeman interaction can also induce  electron transport phenomena~\cite{xu2024efficient}. In this section, we will discuss the impact of the Zeeman interaction to the response.

The incorporation of the Zeeman interaction  brings  an extra  term into  the current density operator \cite{zhong2016gyrotropic}
\begin{eqnarray}\label{57za}
\mathbf{\hat{j}}_{ze}(\mathbf{q})=\frac{ig \mu_B}{V}\mathbf{q}\times \mathbf{S}e^{-i\mathbf{q}\cdot \mathbf{r}}.
\end{eqnarray}
Consequently, we have the following matrix elements
\begin{eqnarray}\label{57zb}
\langle \psi_a(\mathbf{k})|H'_{ze}|\psi_b(\mathbf{k}-\mathbf{q})\rangle&=&\langle u_a(\mathbf{k})|g\mu_B \mathbf{B}(\mathbf{q},\mathbf{\omega})\cdot\mathbf{S}|u_b(\mathbf{k}-\mathbf{q})\rangle,
\end{eqnarray}
\begin{eqnarray}\label{57zc}
\langle \psi_a(\mathbf{k})|H^{'}_{ze}|\psi_b(\mathbf{k}-\mathbf{q}_1)\rangle&=&\langle u_a(\mathbf{k})|g\mu_B \mathbf{B}(\mathbf{q}_1,\mathbf{\omega}_1)\cdot\mathbf{S}|u_b(\mathbf{k}-\mathbf{q}_1)\rangle,
\end{eqnarray}
\begin{eqnarray}\label{57zd}
\langle \psi_a(\mathbf{k}-\mathbf{q}_1))|H^{'}_{ze}|\psi_b(\mathbf{k}-\mathbf{q}_1-\mathbf{q}_2)\rangle&=&\langle u_a(\mathbf{k}-\mathbf{q}_1)|g\mu_B \mathbf{B}(\mathbf{q}_2,\mathbf{\omega}_2)\cdot\mathbf{S}|u_b(\mathbf{k}-\mathbf{q})\rangle,
\end{eqnarray}
\begin{eqnarray}\label{57ze}
\langle \psi_a(\mathbf{k}-\mathbf{q}_1)|\mathbf{\hat{j}}_{ze}(\mathbf{q})|\psi_b(\mathbf{k})\rangle&=&\langle u_a(\mathbf{k}-\mathbf{q}_1)|\frac{ig \mu_B}{V}\mathbf{q}_1\times \mathbf{S}|u_b(\mathbf{k})\rangle,
\end{eqnarray}
\begin{eqnarray}\label{57zf}
\langle \psi_a(\mathbf{k}-\mathbf{q})|\mathbf{\hat{j}}_{ze}(\mathbf{q})|\psi_b(\mathbf{k})\rangle&=&\langle u_a(\mathbf{k}-\mathbf{q})|\frac{ig \mu_B}{V}\mathbf{q}\times \mathbf{S}|u_b(\mathbf{k})\rangle,
\end{eqnarray}
and 
\begin{eqnarray}\label{57zg}
\langle \psi_a(\mathbf{k})|\mathbf{\hat{j}}_{ze}(\mathbf{q})|\psi_b(\mathbf{k})\rangle&=&0.
\end{eqnarray}
By substituting these matrix elements into the expression for the linear response, specifically Eq.~(\ref{13}), we derive a correction that is accurate up to the first order in $\mathbf{q}$
\begin{eqnarray}\label{57zh}
\langle \hat{j}_{\mu}(\mathbf{q},\omega) \rangle_{ze}&=&\frac{eg\mu_B}{\beta V}\sum_{\mathbf{k},\omega_n}\mathrm{Tr}\big[i(\mathbf{q}\times \mathbf{S})_{\mu}
G(\mathbf{k},\omega_n)\hat{v}_{\nu}(\mathbf{k})A_{\nu}(\mathbf{q},\omega)G(\mathbf{k},\omega_n-\omega)\nonumber\\&&-\hat{v}_{\mu}(\mathbf{k})G(\mathbf{k},\omega_n)B_{\rho}(\mathbf{q},\mathbf{\omega})S_{\rho}G(\mathbf{k},\omega_n-\omega)\big]\nonumber\\&=&\frac{eg\mu_B}{V}\sum_{\mathbf{k},a,b}\frac{f_{ab}}{\varepsilon_{ab}-\omega}\big[\langle u_b(\mathbf{k})|S_{\rho}|u_a(\mathbf{k})\rangle\langle u_a(\mathbf{k})|\hat{v}_{\nu}(\mathbf{k})|u_b(\mathbf{k})\rangle \epsilon_{\mu\eta\rho}\epsilon_{\rho\eta\nu}\nonumber\\&&-\langle u_b(\mathbf{k})|\hat{v}_{\mu}(\mathbf{k})|u_a(\mathbf{k})\rangle\langle u_a(\mathbf{k})|S_{\rho}|u_b(\mathbf{k})\rangle\big]B_{\rho}(\mathbf{q},\mathbf{\omega}).
\end{eqnarray}
Taking the limit $\omega\rightarrow 0$, interchanging the indices $a$, and $b$ in the terms involving $f(\varepsilon_b(\mathbf{k}))$
and applying the completeness relation yields
\begin{eqnarray}\label{57zl}
\langle \hat{j}_{\mu}(\mathbf{q},\omega) \rangle_{ze}=\frac{-eg\mu_B}{V}\sum_{\mathbf{k},a}f(\varepsilon_a(\mathbf{k}))\partial_{\mu}\sigma_{a\rho}B_{\rho}(\mathbf{q},\mathbf{\omega}),
\end{eqnarray}
where $\sigma_{a\rho}=\langle u_a(\mathbf{k})|\sigma_{\rho}|u_a(\mathbf{k})\rangle$ and $\partial_{\mu}\sigma_{a\rho}=2\mathrm{Re}\langle u_a(\mathbf{k})|\hat{v}_{\mu}(\mathbf{k})|u_b(\mathbf{k})\rangle\langle u_b(\mathbf{k})|\sigma_{\rho}|u_a(\mathbf{k})\rangle/\varepsilon_{ab}$. 
It is worth noting that the term $\mathbf{\hat{j}}_{ze}(\mathbf{q})$ and  the Zeeman interaction term in the current, namely Eq.~(\ref{57zh}), yield identical contribution.
 
Now, let us consider the contribution of the Zeeman interaction to the second-order response in the SHG case. By substituting the matrix elements, namely Eqs.~(\ref{57zb})-(\ref{57zg}), into the expressions for the nonlinear response, i.e., Eqs.~(\ref{25})-(\ref{28}), We obtain a correction that is accurate to first order in $\mathbf{q}_0$
\begin{eqnarray}\label{26s}
\langle \hat{j}_{\mu}(2\mathbf{q}_0,2\omega_0) \rangle_{ze}&=&\frac{-2e^2 g\mu_B}{\beta V}\sum_{\mathbf{k},\omega_n}\mathrm{Tr}\bigg[\partial_{\mu}\hat{v}_{\gamma}(\mathbf{k})G(\mathbf{k},\omega_n)S_{\rho}G(\mathbf{k},\omega_n-\omega_0)A_{\gamma}(\mathbf{q}_0,\omega_0)B_{\rho}(\mathbf{q}_0,\omega_0)\nonumber\\&&-i(\mathbf{q}_0\times \mathbf{S})_{\mu}G(\mathbf{k},\omega_n)\partial_{\nu}\hat{v}_{\gamma}(\mathbf{k})G(\mathbf{k},\omega_n-2\omega_0)A_{\gamma}(\mathbf{q}_0,\omega_0)A_{\nu}(\mathbf{q}_0,\omega_0)\nonumber\\&&+\hat{v}_{\mu}(\mathbf{k})G(\mathbf{k},\omega_n)\hat{v}_{\gamma}(\mathbf{k})G(\mathbf{k},\omega_n-\omega_0) S_{\rho}G(\mathbf{k},\omega_n-2\omega_0)A_{\gamma}(\mathbf{q}_0,\omega_0)B_{\rho}(\mathbf{q}_0,\omega_0)\nonumber\\&&+\hat{v}_{\mu}(\mathbf{k})G(\mathbf{k},\omega_n)S_{\rho}G(\mathbf{k},\omega_n-\omega_0) \hat{v}_{\gamma}(\mathbf{k})G(\mathbf{k},\omega_n-2\omega_0)A_{\gamma}(\mathbf{q}_0,\omega_0)B_{\rho}(\mathbf{q}_0,\omega_0)\nonumber\\&&-i(\mathbf{q}_0\times \mathbf{S})_{\mu}G(\mathbf{k},\omega_n)\hat{v}_{\gamma}(\mathbf{k})G(\mathbf{k},\omega_n-\omega_0) \hat{v}_{\nu}(\mathbf{k})G(\mathbf{k},\omega_n-2\omega_0)A_{\gamma}(\mathbf{q}_0,\omega_0)A_{\nu}(\mathbf{q}_0,\omega_0)\nonumber\\&&-i(\mathbf{q}_0\times \mathbf{S})_{\mu}G(\mathbf{k},\omega_n)\hat{v}_{\nu}(\mathbf{k})G(\mathbf{k},\omega_n-\omega_0) \hat{v}_{\gamma}(\mathbf{k})G(\mathbf{k},\omega_n-2\omega_0)A_{\gamma}(\mathbf{q}_0,\omega_0)A_{\nu}(\mathbf{q}_0,\omega_0)\nonumber\\&=&
\frac{-2e^2 g\mu_B}{V}\sum_{\mathbf{k},a,b}\bigg[\frac{1}{\varepsilon_{ab}-\omega_0}\bigg(\frac{f_{ac}}{\varepsilon_{ac}-2\omega_0}-\frac{f_{bc}}{\varepsilon_{bc}-\omega_0}\bigg)\nonumber\\&&\times\big[\langle u_c(\mathbf{k})|\hat{v}_{\mu}(\mathbf{k})|u_a(\mathbf{k})\rangle
\langle u_a(\mathbf{k})|\hat{v}_{\gamma}(\mathbf{k})|u_b(\mathbf{k})\rangle\langle u_b(\mathbf{k})|S_{\rho}|u_c(\mathbf{k})\rangle\nonumber\\&&+\langle u_c(\mathbf{k})|\hat{v}_{\mu}(\mathbf{k})|u_a(\mathbf{k})\rangle\langle u_a(\mathbf{k})|S_{\rho}(\mathbf{k})|u_b(\mathbf{k})\rangle\langle u_b(\mathbf{k})|\hat{v}_{\gamma}(\mathbf{k})|u_c(\mathbf{k})\rangle\nonumber\\&&+\langle u_c(\mathbf{k})|S_{\rho}(\mathbf{k})|u_a(\mathbf{k})\rangle\langle u_a(\mathbf{k})|\hat{v}_{\gamma}(\mathbf{k})|u_b(\mathbf{k})\rangle\langle u_b(\mathbf{k})|\hat{v}_{\mu}(\mathbf{k})|u_c(\mathbf{k})\rangle\nonumber\\&&+\langle u_c(\mathbf{k})|S_{\rho}(\mathbf{k})|u_a(\mathbf{k})\rangle\langle u_a(\mathbf{k})|\hat{v}_{\mu}(\mathbf{k})|u_b(\mathbf{k})\rangle\langle u_b(\mathbf{k})|\hat{v}_{\gamma}(\mathbf{k})|u_c(\mathbf{k})\rangle\big]\nonumber\\&&+\frac{f_{ab}}{\varepsilon_{ab}-\omega_0}\langle u_b(\mathbf{k})|\partial_{\mu}\hat{v}_{\gamma}(\mathbf{k})|u_a(\mathbf{k})\rangle\langle u_a(\mathbf{k})|S_{\rho}(\mathbf{k})|u_b(\mathbf{k})\rangle\nonumber\\&&+\frac{f_{ab}}{\varepsilon_{ab}-2\omega_0}\langle u_b(\mathbf{k})|S_{\rho}|u_a(\mathbf{k})\rangle\langle u_a(\mathbf{k})|\partial_{\mu}\hat{v}_{\gamma}(\mathbf{k})|u_b(\mathbf{k})\rangle\bigg]A_{\gamma}(\mathbf{q}_0,\omega_0)B_{\rho}(\mathbf{q}_0,\omega_0).
\end{eqnarray}
For a two-band system, when the indices  $a, b$, and $c$ correspond to the same band, the above result will be zero. Therefore,the permissible combinations of indices are  $a,b,a;a,b,b;a,a,b$, respectively. Subsequently, by interchanging  the indices  $a$ and $b$ in terms involving  $f(\varepsilon_b(\mathbf{k}))$, we can derive the correction to the zeroth and first-order response coefficients as 
\begin{eqnarray}\label{38an}
\alpha^{(0)ze}_{\mu\gamma\rho}=\frac{-2e^2 g\mu_B}{i\omega_0 V}\sum_{\mathbf{k},a}f(\varepsilon_a(\mathbf{k}))\partial_{\mu}\partial_{\gamma}\sigma_{a\rho},
\end{eqnarray}
and 
\begin{eqnarray}\label{38am}
\alpha^{(1)ze}_{\mu\gamma\rho}=\frac{e^2 g\mu_B}{ V}\sum_{\mathbf{k},a}f(\varepsilon_a(\mathbf{k}))\big(2\partial_{\mu}Z^{ab}_{\gamma\rho}-\partial_{\gamma}Z^{ab}_{\mu\rho}\big),
\end{eqnarray}
where $Z^{ab}_{\gamma\rho}=-2\mathrm{Im}\langle u_a(\mathbf{k})|\hat{v}_{\gamma}(\mathbf{k})|u_b(\mathbf{k})\rangle\langle u_b(\mathbf{k})|\sigma_{\rho}|u_a(\mathbf{k})\rangle/\varepsilon^2_{ab}$ is a quantity analogous to  Berry curvature~\cite{xiang2025intrinsic}.
 The interaction Hamiltonian discussed in this section is linked  to  the spin magnetic moment, whereas the interaction Hamiltonian explored in the previous sections is primarily  associated with the orbital magnetic moment. Since the orbital magnetic moment is considerably  larger than the spin magnetic moment, the results obtained in this section are expected to be comparatively smaller than those derived in the earlier sections.

\section{Concluding remarks}
In this paper, we utilize  Green's function to investigate  the electromagnetic response in systems characterized by a one-body Hamiltonian.  When two-body interactions, such as electron-electron interactions, are incorporated, Green's function remains  a powerful tool. In fact, by differentiating the Green's function with respect to time or employing  perturbative expansion technique from quantum field theory \cite{mahan2013many}, one can derive the exact Green's function as well as  the self-energy for systems with two-body interactions. 

The nonlinear response can also be studied using lesser, retarded, and advanced Green's functions~\cite{xu2022abnormal}. This approach is particularly advantageous for investigating responses in systems with impurities. However, for clean systems, it is equivalent to the Matsubara Green's function method that we have employed in this paper.

Here, for simplicity, we focus on the response up to the second order. Following the same methodology, the third-order response to the vector potential can be directly derived.
Our formalism is also applicable for analyzing inhomogeneous electromagnetic responses ~\cite{zhang2019tunable,kozii2021intrinsic,zhang2022geometric,gassner2023regularized,mckay2024charge}.
 Furthermore, the method presented in this paper can be employed to derive unconventional responses, such as thermal responses ~\cite{wang2022quantum}.

\section*{Acknowledgements}
A. Z. acknowledges the support from Shanghai Magnolia Talent Plan  Youth Project and Shanghai Normal University for initial start-up funding (Grant No. 307-AF0102-25-005323).

\section*{Data availability}
The data are available from the corresponding author
upon reasonable request.

\bibliographystyle{unsrt}
\bibliography{ref}

\end{document}